\documentclass[journal]{IEEEtran}
\IEEEoverridecommandlockouts
\usepackage{cite}
\usepackage{amsmath,amssymb,amsfonts}
\usepackage{graphicx}
\usepackage{textcomp}
\usepackage{xcolor}
\usepackage{array}
\usepackage[ruled,vlined,linesnumbered]{algorithm2e}
\usepackage[absolute,overlay]{textpos}  
\usepackage[compatibility=false]{caption}
\usepackage{subcaption}
\usepackage{ltablex}
\usepackage{pdflscape}  
\usepackage{afterpage}  
\usepackage{makecell}   
\usepackage{booktabs}
\usepackage{cuted}
\usepackage{color}
\newcolumntype{C}[1]{>{\centering\arraybackslash}m{#1}}

\makeatletter
\let\orig@subref\subref 
\renewcommand{\subref}[1]{\thefigure\orig@subref{#1}} 
\makeatother

\allowdisplaybreaks[4]

\def\BibTeX{{\rm B\kern-.05em{\sc i\kern-.025em b}\kern-.08em
    T\kern-.1667em\lower.7ex\hbox{E}\kern-.125emX}}
\begin{document}

\title{Age of Information Optimization in Laser-charged UAV-assisted IoT Networks: A Multi-agent Deep Reinforcement Learning Method\\}

\author{
\author{Geng~Sun,
        Likun~Zhang, 
        Jiahui Li,
        Jing Wu,
        Jiacheng Wang,
        Zemin Sun,
        Changyuan Zhao,\\
        Victor C.M. Leung, \IEEEmembership{Life Fellow,~IEEE}

    \thanks{This study is supported in part by the National Natural Science Foundation of China (62272194, 62471200), in part by the Science and Technology Development Plan Project of Jilin Province (20250101027JJ), in part by the Postdoctoral Fellowship Program of China Postdoctoral Science Foundation (GZC20240592), in part by China Postdoctoral Science Foundation General Fund (2024M761123), and in part by the Scientific Research Project of Jilin Provincial Department of Education (JJKH20250117KJ).
    
    \textit{(Corresponding author: Jing Wu.)}
   
    \par Geng Sun is with the College of Computer Science and Technology, Key Laboratory of Symbolic Computation and Knowledge Engineering of Ministry of Education, Jilin University, Changchun 130012, China, and also with the College of Computing and Data Science, Nanyang Technological University, Singapore 639798 (e-mail: sungeng@jlu.edu.cn).

    \par Likun Zhang, Jiahui Li, Jing Wu, and Zemin Sun are with the College of Computer Science and Technology, Jilin University, Changchun 130012, China (e-mails: zhanglk23@mails.jlu.edu.cn, lijiahui@jlu.edu.cn, wujing@jlu.edu.cn, sunzemin@jlu.edu.cn).
    
    \par Jiacheng Wang and Changyuan Zhao are with the College of Computing and Data Science, Nanyang Technological University, Singapore 639798 (e-mails: jiacheng.wang@ntu.edu.sg, zhao0441@e.ntu.edu.sg).
    
    \par Victor C. M. Leung is with the College of Computer Science and Software Engineering, Shenzhen University, Shenzhen 518060, China, and also with the Department of Electrical and Computer Engineering, University of British Columbia, Vancouver, BC V6T 1Z4, Canada (e-mail: vleung@ieee.org).
    }

}

}

\maketitle

\begin{abstract}
The integration of unmanned aerial vehicles (UAVs) with Internet of Things (IoT) networks offers promising solutions for efficient data collection. However, the limited energy capacity of UAVs remains a significant challenge. In this case, laser beam directors (LBDs) have emerged as an effective technology for wireless charging of UAVs during operation, thereby enabling sustained data collection without frequent returns to charging stations (CSs). In this work, we investigate the age of information (AoI) optimization in LBD-powered UAV-assisted IoT networks, where multiple UAVs collect data from distributed IoTs while being recharged by laser beams. We formulate a joint optimization problem that aims to minimize the peak AoI while determining optimal UAV trajectories and laser charging strategies. This problem is particularly challenging due to its non-convex nature, complex temporal dependencies, and the need to balance data collection efficiency with energy consumption constraints. To address these challenges, we propose a novel multi-agent proximal policy optimization with temporal memory and multi-agent coordination (MAPPO-TM) framework. Specifically, MAPPO-TM incorporates temporal memory mechanisms to capture the dynamic nature of UAV operations and facilitates effective coordination among multiple UAVs through decentralized learning while considering global system objectives. Simulation results demonstrate that the proposed MAPPO-TM algorithm outperforms conventional approaches in terms of peak AoI minimization and energy efficiency. Ideally, the proposed algorithm achieves up to 15.1\% reduction in peak AoI compared to conventional multi-agent deep reinforcement learning (MADRL) methods.
\end{abstract}


\section{Introduction}

\par The rapid development of unmanned aerial vehicles (UAVs) and the Internet of Things (IoTs) has sparked significant interest in their integration to create efficient and scalable systems for data collection, energy replenishment, and communication in various applications \cite{liu2022fair, liu2022joint}. This integration has emerged as a promising paradigm to address fundamental challenges in conventional IoT infrastructures, particularly regarding energy efficiency, communication range, and data collection capabilities. UAVs, characterized by their high mobility, flexibility, and ability to establish line-of-sight (LoS) communication links, offer a viable solution to the limitations of conventional IoT networks where battery life and communication range are often constrained \cite{lahmeri2022charging}.

\par However, one of the critical challenges in UAV-assisted IoT networks is the limited energy capacity of UAVs. In particular, the substantial propulsion energy consumption during flight operations, coupled with the energy requirements for communication and data processing, significantly restricts the operational duration of UAVs \cite{panahi2023reliable, liu2018age}. Conventional approaches relying on fixed charging stations (CSs), while useful in some scenarios, introduce additional complexity and operational overhead, particularly when UAVs are deployed over large areas or in environments without accessible infrastructure \cite{panahi2023reliable, song2024energy}.

\par To overcome these limitations, recent advancements have proposed integrating UAVs with wireless charging technologies. Laser charging, in particular, has gained significant attention due to its ability to deliver concentrated energy over long distances \cite{benmad2021data, zhang2021age}, thus allowing UAVs to recharge while still performing their data collection tasks \cite{ranjha2021urllc}. Laser charging systems, such as those utilizing laser beam directors (LBDs) and high-power lasers, offer several advantages over traditional radio frequency (RF)-based charging, particularly in scenarios where UAVs must cover extensive areas or operate over long distances without returning to a fixed CS \cite{ahmadi2025advancements, zhu2022aerial}. However, deploying laser charging systems introduces its own set of challenges, including the need for accurate beam control  \cite{zeng2019energy}, optimal energy transfer efficiency \cite{wu2020load}, and coordination between charging and data collection operations \cite{lu2024optimal}. In particular, the performance metric of the data collection operations may have trade-offs with the charging efficiency of LBDs, which pose more requirements for designing the LBD-powered UAV systems. 

\par In this work, we aim to investigate age of information (AoI) optimization in the LBD-powered UAV systems, which is a critical metric for evaluating the timeliness and freshness of information in IoT networks \cite{liu2018age, hu2020aoi}. This metric is particularly important in scenarios requiring real-time monitoring and decision-making, such as emergency response, industrial automation, and smart city applications. Minimizing AoI ensures that the collected data accurately represents the current state of the monitored environment, thereby enhancing the effectiveness of IoT-based systems \cite{amodu2024deep, long2024aoi}. As aforementioned, the integration of AoI optimization with LBD-powered UAV systems presents a complex challenge that requires sophisticated approaches to balance information freshness with energy efficiency \cite{zhang2021age, guo2024aoi}.

\par Traditional optimization approaches such as convex optimization and mathematical programming have been widely applied to UAV trajectory planning and resource allocation problems \cite{shen2023convex, Zhang2024Interactive, Zhang2024GenerativeAI}. While these methods provide optimal solutions under specific conditions, they often struggle with the real-time adaptability and dynamic nature required in LBD-powered UAV systems. Particularly, the precise beam tracking and continuous adjustment needed for effective laser charging demand solutions that can respond instantaneously to changing conditions, which conventional optimization techniques cannot adequately address \cite{chen2024survey}. In this case, deep reinforcement learning (DRL) approaches have gained significant attention due to their ability to adapt to dynamic environments and handle complex decision-making processes \cite{yi2020deep}. Various DRL techniques, including deep Q-network (DQN) \cite{li2019board}, deep deterministic policy gradient (DDPG) \cite{miao2024utility}, and proximal policy optimization (PPO) \cite{ghomri2024drl}, have demonstrated promising results in optimizing UAV trajectories and resource allocation. However, these centralized DRL methods often fall short when dealing with the distributed characteristics of one-to-many charging scenarios common in LBD-powered UAV systems, where multiple UAVs need to coordinate their actions while operating in different regions \cite{li2022applications}.

\par Thus, in this work, we first conduct a comprehensive survey of the current state of LBD-powered UAV systems, identifying the key challenges and limitations of existing approaches. Based on this analysis, we seek to propose a novel distributed DRL framework that addresses the specific requirements of LBD-powered UAV-assisted IoT data collection systems, thereby enabling effective coordination among multiple UAVs while accounting for the temporal dependencies and spatial constraints inherent in these systems. The main contributions of this work are summarized as follows:

\begin{itemize}
\item \textit{LBD-powered UAV Data Collection System:} We consider a UAV-assisted IoT network where UAVs collect data from IoTs while being recharged by laser beams from LBDs. In such systems, the UAVs operate in both charging and non-charging areas, collecting data from IoTs and optimizing the energy usage during the data collection process. This system is capable of providing efficient and sustainable data collection solutions in large-scale and energy-constrained environments \cite{alsadie2024efficient}.

\item \textit{Joint Optimization of UAV Trajectory and Laser Charging:} In the considered system, we formulate a joint optimization problem that aims to minimize the peak AoI in the network by optimizing the UAV trajectories and charging strategies. This optimization actually needs to balance the data collection efficiency, energy consumption during flight and communication, and the effective use of laser charging for energy replenishment \cite{liau2024laser}. Thus, this problem is challenging due to its dynamic nature and the need to adapt to varying energy levels, IoT locations, and UAV positions in real-time.

\item \textit{MADRL-based Approach:} To solve the formulated optimization problem, we propose an MADRL-based approach, namely multi-agent proximal policy optimization with temporal memory and multi-agent coordination (MAPPO-TM). In MAPPO-TM, the incorporated temporal memory captures the dynamic nature of UAV trajectories and energy consumption while enabling coordination among multiple UAVs. This approach efficiently learns the optimal strategies through decentralized learning while considering the global system objectives.

\item \textit{Simulation and Performance Evaluation:} Simulation results demonstrate that the proposed MAPPO-TM algorithm outperforms conventional approaches in terms of peak AoI minimization and UAV energy efficiency. Thus, the algorithm effectively coordinates multiple UAVs, reduces the AoI, and optimizes the UAV trajectories and laser charging strategies. 

\end{itemize}

\par The rest of this paper is organized as follows. Section~\ref{sec: related work} reviews the related works. Section~\ref{sec: system model} presents the models and problem formulation. Section~\ref{sec: MADRL-based approach} introduces the proposed MAPPO-TM algorithm. In Section~\ref{sec: simulation results}, we present the simulation results. Finally, the paper is concluded in Section~\ref{sec: conclusion}.




{\footnotesize
\begin{table*}[t]
  \centering
  \caption{{\\ \textsc{Comparisons between Related Works with This Work.}}}
  \label{comparison}
  \setlength{\tabcolsep}{2pt}
  \renewcommand{\arraystretch}{1}
  \begin{tabular}{|C{1cm}|C{1.2cm}|C{3.5cm}|C{3.5cm}|C{3.5cm}|C{4.5cm}|}

\hline
\textbf{Related works} & \textbf{Number of UAVs} & \textbf{Communication scenarios} & \textbf{Optimization variables} & \textbf{Network optimization metrics} & \textbf{Optimization methods} \\
\hline

[30] & Single & UAV-aided relay communication for edge computing & DBS placement, bandwidth, power and UAV trajectory & DBS service time and task completion time for all UEs & Iterative algorithm and placement algorithm based on counting sort \\
\hline
[6] & Single & UAV-assisted MEC with HAP wireless charging & UAV flight trajectories & Energy efficiency and number of computation tasks collected by the UAV & Multi-objective reinforcement learning with trace-based experience replay \\
\hline
[9] & Single & Laser-powered UAV relay system for URLLC connectivity between source node and ground station & UAV trajectory, blocklength allocation, power control, and EH & Total decoding error rate, energy efficiency, and UAV energy consumption & Perturbation-based iterative method with divide-and-conquer approach \\
\hline
[29] & Single & Data gathering from ground IoTs with laser charging from HAPs & UAV hovering positions and charging energies at each position & Total task completion time (including data collection and charging time) & BCD approach, SCA and dynamic programming\\
\hline
[43] & Single & Air-ground coordinated MEC system with laser-powered UAV serving as both MEC server and relay for ground access point & Trajectory of UAV, computation task allocation between UAV and access point, and EH time allocation & Long-term average energy consumption of UAV & Two-step alternating optimization algorithm that combines linear programming for task and time allocation with DDPG for trajectory design \\
\hline
[46] & Multiple & UAV-assisted wireless powered MEC for metaverse applications & Charging time, computation tasks scheduling, UAV trajectory design & Computation efficiency & Two-stage alternating optimization algorithm based on multi-task DRL \\
\hline
[47] & Single & UAV-based data collection and wireless charging for IoTs & 3D trajectory of UAV, IoT scheduling & Residual energy while satisfying IoT requirements & SCA and BCD algorithm \\
\hline
[5] & Single & UAV-enabled wireless sensor networks for data collection & Flight trajectory of UAV & AoI minimization and average AoI minimization & Dynamic programming and genetic algorithm \\
\hline
[17] & Multiple & Multi-UAV-assisted wireless backscatter networks for sensing data collection & Access control of ground users, beamforming and trajectory planning of UAVs & Long-term time-averaged AoI & Lyapunov optimization, BCD, soft actor-critic algorithm \\
\hline
[56] & Single & Aerial-ground collaborative MEC & UAV flight paths, task offloading ratios & Total AoI of ground devices and total energy consumption of UAV & Multi-objective learning algorithm based on PPO \\
\hline
[57] & Single & Visible light communication-based vehicles-to-everything communication with cluster-based architecture & UAV flight paths, task offloading ratios & Energy efficiency and AoI & MADRL with TD3 \\
\hline
[60] & Single & UAV collecting data from SNs distributed across multiple islands & Transmit power of SNs, clustering of islands, UAV flight trajectory & Long-term average AoI & Clustering-based dynamic adjustment of the shortest path algorithm \\
\hline
[67] & Multiple & UAV-aided information gathering from multiple sources over unreliable wireless channels & Transmission scheduling decisions & Throughput maximization with per-source AoI guarantee & Confidence bound based oracle stationary randomized sampling algorithm \\
\hline
[79] & Multiple & UAV-assisted MEC system & Task offloading decisions, computation resource allocation, UAV trajectory control & Task completion delay, UAV energy consumption, number of offloaded tasks & JTORATC approach using distributed splitting, threshold rounding, Karush-Kuhn-Tucker method, and SCA \\
\hline
[23] & Single & UAV collects status update packets from distributed sensors & UAV flight trajectory, transmission scheduling & Weighted sum of AoI & DRL method \\
\hline
[90] & Multiple & NOMA-enabled multi-UAV collaborative caching network & Caching decision, 3D trajectory planning, power allocation, subchannel reusing & System content retrieving delay & MAPPO and matching-DRL solution \\
\hline
Our work & Multiple & UAVs collect data from IoTs while recharged by LBDs & UAV trajectories and laser charging strategies & Peak AoI minimization & MAPPO-TM algorithm \\
\hline

\end{tabular}
\end{table*}
}

\section{Related Work}
\label{sec: related work}

\par In this work, we aim to charge UAVs with lasers and optimize the flight trajectory of UAVs to better execute charging strategies and efficiently collect data in IoT networks. In the following, we will review some key related works to illustrate the novelty of our research. 
The comparisons between these related works and our work are shown in Table \ref{comparison}.

\subsection{Laser-powered UAVs Assisted Communications}

\subsubsection{Laser-powered UAV System Architectures and Fundamentals}

\par Many existing works explored laser-powered UAV system architectures for data collection in IoT networks. For instance, the authors in \cite{lahmeri2022charging} proposed a basic IoT structure integrating UAVs, IoTs, and LBDs, establishing the framework where UAVs collect data while LBDs provide energy. The authors in \cite{ahmadi2025advancements} explored the technical fundamentals of laser charging and classified optical wireless power transfer (OWPT) systems into laser power transfer (LPT) and LED-based OWPT, highlighting the advantages of LPT for high-density, long-distance energy transfer in applications like autonomous UAVs. Moreover, the authors in \cite{zhu2022aerial} categorized UAVs into charging UAVs (CUAVs) and mission UAVs (MUAVs) to enable aerial refueling without mission interruption. The authors in \cite{Liu2022JointLaser} developed a laser charging framework employing a drone-mounted base station (DBS) that concurrently delivers services to ground-based user equipments (UEs) while harvesting energy transmitted from a laser CS mounted on the macro base station (MBS). Within their system architecture, both the MBS and DBS incorporate computational servers, enabling UEs to offload computational tasks to either station, while the DBS is precisely positioned at optimal locations to enhance uplink communications and computing services for ground UEs during each operational time slot. In addition, the authors in \cite{lahmeri2022laser} investigated various deployment strategies for such systems, analyzing coverage capabilities under different laser charging conditions and optical turbulence through stochastic geometry. The authors in \cite{nguyen2023dynamic} proposed a dynamic OWPT system with overhead facilities housing laser transmitters and tracking cameras for continuous charging of moving vehicles. The authors in \cite{Zhang2023Privacy} examined the integration of laser-beamed wireless power transfer into high-altitude platform (HAP)-aided multiaccess edge computing systems serving HAP-connected aerial user equipments. By discretizing the three-dimensional (3D) coverage space of the HAP, they established a sophisticated multitier tile grid-based spatial structure that furnished aerial location options in the form of tile grids for effective laser charging of aerial user equipments. The authors in \cite{Cheng2021Intelligent} formulated an energy-constrained UAV-aided mobile edge-cloud continuum framework wherein offloaded tasks from ground IoTs can be cooperatively executed by UAVs functioning as edge servers and cloud servers connected to a ground base station (GBS) that serves as an access point. In their particular framework implementation, UAVs powered by laser beams transmitted from the GBS subsequently provide wireless charging capabilities to IoTs. However, most existing studies did not fully address the integration of continuous charging areas with non-charging areas, which significantly limited their practical application in real-world scenarios where UAVs should operate across diverse operational zones.

\subsubsection{Resource Allocation and Trajectory Optimization in Laser-powered System}

\par Many existing works have focused on optimizing resource allocation and flight trajectories in laser-powered systems. For instance, the authors in \cite{panahi2023reliable} proposed a cost-aware UAV deployment strategy to ensure high-quality communication and energy links between UAVs and ground users. The authors in \cite{song2024energy} investigated energy-efficient trajectory optimization with wireless charging for UAV-assisted mobile edge computing (MEC). Moreover, the authors in \cite{ranjha2021urllc} studied joint resource allocation, trajectory design, and energy harvesting (EH) to achieve ultra-reliable and low-latency communication (URLLC) in laser-powered UAV relay systems. The authors in \cite{lu2024optimal} proposed a multimodal charging system with a two-layer model considering both inductive wireless power transfer and laser power scheduling. The authors in \cite{liau2024laser} introduced the minimum completion time trajectory and charging optimization algorithm, optimizing hovering positions and charging energies using block coordinate descent (BCD). In addition, the authors in \cite{du2024dynamic} presented solutions to optimize energy efficiency of UAV relaying in IoT systems through trajectory planning and bandwidth allocation. The authors in \cite{li2023data} developed an improved clustering algorithm to determine optimal visiting order and entry points for IoT clusters, maximizing system energy efficiency. These works have significantly advanced trajectory and resource optimization in laser-powered UAV systems. The authors in \cite{Bashir2023Energy} examined the simultaneous lightwave information and power transfer scheme for laser-powered decode-and-forward UAV relays functioning within an optical wireless backhaul. Their primary objective centered on determining the optimal allocation of received beam energy across the decoding circuit, transmitting circuit, and rotor block of the relay to maximize quality-of-service metrics including achievable data rate, outage probability, and error probability. The authors in \cite{Liu2021Laser} developed a laser charging-enabled DBS framework where a ground-based laser CS continuously delivers energy to an aerial quadrotor DBS providing communication services to multiple users. They formulated a comprehensive optimization problem addressing joint power and bandwidth assignment along with laser charging-enabled DBS placement to simultaneously maximize both flight duration and communication data rates. The authors in \cite{Park2022Joint} explored a laser-charged UAV relaying network wherein a rotary-wing UAV serves as a data relay between a GBS and UE while simultaneously receiving power from a dedicated CS via laser beam transmission. They addressed the critical optimization challenge of minimizing CS power consumption while ensuring the minimum data requirements of UE, consequently developing an algorithm that jointly optimizes UAV trajectory and charging power allocation. However, the aforementioned studies typically employed conventional optimization methods that struggled with real-time adaptability required for dynamic environments, limiting their effectiveness in scenarios with rapidly changing conditions or incomplete environmental information.

\subsubsection{Various Applications of Laser-powered Systems}

\par Many existing studies have investigated specific applications of laser-powered UAVs in various domains. For example, the authors in \cite{ma2024green} employed green energy-powered base stations with laser chargers to extend UAV uptime for wireless charging and data backhauling in wireless rechargeable sensor networks. The authors in \cite{abdelhady2024operation} investigated maximizing harvested data in laser-powered UAV-supported IoT deployments, enabling battery-free IoTs to establish communication links via bistatic backscattering. Moreover, the authors in \cite{zhang2024resource} and \cite{wang2024air} explored UAV-assisted edge computing, with the latter proposing an air-ground coordinated MEC system where a laser-powered UAV served as both an MEC server and relay. The authors in \cite{singh2023laser} investigated multi-UAV systems with full-duplex mobile users under URLLC constraints. In addition, the authors in \cite{singh2022performance} presented a UAV-assisted multiuser network using laser charging and simultaneous wireless information and power transfer. The authors in \cite{wang2023wireless} proposed a two-stage optimization algorithm based on multi-task DRL for laser charging in wireless-powered metaverse scenarios. The authors in \cite{Fu2021Joint} investigated a solar-powered UAV system where the UAV simultaneously collects data from IoTs on the ground and charges them utilizing laser charging technology. Their objective focused on maximizing the residual energy of UAV while fulfilling IoT requirements through joint optimization of the 3D trajectory of UAV and IoT scheduling protocols. The authors in \cite{Goel2023Backscatter} examined the modeling of data collection from backscatter nodes utilizing UAVs that maintain sustainable operations through wireless energy transfer from laser based CSs. The authors in \cite{betalo2025dynamic} introduced an energy-efficient laser-charged UAV-enabled rechargeable wireless sensor network environment wherein UAVs, energized by laser beams transmitted from ground-based stations, deliver services, gather data, and transfer energy to sensor nodes (SNs).  Their research formulated a sophisticated joint optimization problem encompassing power allocation, dynamic charging strategy, and path planning with the dual objectives of minimizing task completion time and SN death time. However, most of the above works did not adequately address the challenge of balancing data collection efficiency with energy management across charging and non-charging areas, limiting their comprehensive application in complex IoT networks.

\par Different from these existing works, this work uniquely designs a laser-powered UAV system that effectively integrates operations across both charging and non-charging zones, employs advanced DRL methods for adaptive trajectory optimization in dynamic environments, and comprehensively addresses the challenge of balancing data collection efficiency with energy management in complex IoT networks.

\subsection{Optimizations of AoI in IoT Networks}

\subsubsection{UAV Trajectory Planning for AoI Optimization}

\par Many existing works have studied UAV path planning to minimize AoI in IoT networks. For instance, the authors in \cite{liu2018age} studied the age-optimal trajectory planning problem in UAV-enabled IoT networks, designing optimal trajectories to minimize both the age of the oldest sensed information and the average AoI of all IoTs. The authors in \cite{benmad2021data} explored data collection in UAV-assisted IoT networks powered by harvested energy, aiming to minimize the mission total time while ensuring each IoT receives required energy and transfers its sensed data. Moreover, the authors in \cite{zhang2021age} studied average AoI optimization by optimizing UAV trajectory in energy recharging networks, where the UAV collects data from IoTs and is replenished by ground chargers. The authors in \cite{hu2020aoi} investigated average AoI minimization based on UAV trajectory and time allocation for EH and data collection, where the UAV serves as both mobile data collector and charger for IoTs. The authors in \cite{amodu2024deep} presented a review of UAV-aided data collection focusing on DRL approaches to minimize AoI. In addition, the authors in \cite{long2024aoi} formulated a multi-stage stochastic optimization to minimize long-term time-averaged AoI by jointly optimizing access control, beamforming, and trajectory planning for multiple UAVs. The authors in \cite{zhang2025aoi} constructed a space-air-ground integrated network with satellites, HAPs, UAVs, and terrestrial IoTs, minimizing system AoI through UAV trajectory design and network configuration. The authors in \cite{Zhu2023UAV} developed an optimization framework aimed at minimizing the total AoI of data collected by UAVs from ground IoT networks. Recognizing that the total AoI depends critically on both UAV flight time and data collection duration at hovering points, they conducted joint optimization of hovering point selection and the sequential visiting order to these locations. The authors in \cite{Dang2022AoI} introduced a sophisticated DRL-based proactive UAV trajectory planning algorithm capable of autonomously adjusting flight policies in response to channel variations while balancing the trade-off between energy transmission and data collection, ultimately achieving optimal system-level AoI under dynamic channel conditions. The authors in \cite{Xiao2024Joint} addressed the complexity of multiple UAV operations by formulating a joint multi-UAV trajectory planning and data collection problem as a mixed integer nonlinear programming model, with the dual objectives of minimizing both AoI and energy consumption. However, most existing studies did not fundamentally solved the energy issues associated with extended UAV operation, which impacted their ability to achieve sustained information freshness in long-duration missions.

\subsubsection{Trade-offs between Energy Efficiency and AoI}

\par Many existing works have addressed the crucial balance between energy efficiency and information freshness in networked systems. For example, the authors in \cite{wang2024aoi} incorporated AoI as a metric to ensure information freshness and designed an AoI-aware energy efficiency resource allocation scheme for satellite-based IoT networks. The authors in \cite{cao2024risk} tackled the problem of selecting the optimal number of connections that is both AoI-optimal and energy-efficient by introducing an energy efficiency-peak AoI ratio to enable a trade-off between AoI and energy consumption. Moreover, the authors in \cite{song2024aoi} studied the AoI and energy trade-off in an aerial-ground collaborative MEC system, formulating a multi-objective optimization problem to simultaneously minimize total AoI and energy consumption by optimizing flight paths and task offloading ratios. The authors in \cite{azizi2024efficient} explored energy efficiency and AoI awareness in a cluster-based visible light communication vehicles-to-everything system, evaluating the impact of vehicle numbers on energy efficiency and AoI. The authors in \cite{sun2024aoi} studied AoI optimization in information-gathering wireless networks where sources are equipped with batteries harvesting ambient energy, analyzing how energy arrival patterns and transmission policies influence average AoI. In addition, the authors in \cite{chen2025average} investigated average AoI optimization in wireless-powered networks with directional charging, proposing an AoI-aware periodical charging scheduling algorithm. The authors in \cite{liu2024aoi} investigated the long-term average AoI-minimal problem in a UAV-assisted wireless-powered communication network spanning multiple islands. The authors in \cite{Yi2023Multitask} examined the challenge of rechargeable-UAV-aided timely data collection in IoT networks, wherein the UAV gathers status updates from multiple sensors while maintaining its energy level above a required threshold through recharging. To balance information freshness against energy consumption, they developed a Markov decision process framework designed to minimize the weighted sum of average total AoI and average recharging price. The authors in \cite{Zhang2024AoI} investigated the AoI and energy trade-off within a system utilizing a UAV for data collection across multiple IoT nodes. To thoroughly analyze the interplay between AoI and energy consumption, they conducted joint optimization of collection time, UAV trajectory, and time slot duration. The authors in \cite{Zhao2024Safe} explored task offloading challenges in a UAV-aided wireless powered edge computing system, emphasizing information freshness enhancement while maintaining UAV energy safety. To achieve minimum average AoI, they proposed a comprehensive joint optimization approach encompassing ground device wireless charging power, UAV flight trajectory, and offloading decisions. However, although most of the above works made significant contributions to understanding energy-AoI trade-offs, they did not fundamentally solve the energy issues, which limited their effectiveness in scenarios requiring extended operation periods.

\subsubsection{AoI in Specialized Networks and Applications}

\par Many existing studies have explored AoI optimization in diverse application domains with specific requirements. For instance, the authors in \cite{guo2024aoi} introduced a theoretical framework for optimizing second-order behaviors of wireless networks, making it well-suited for modeling AoI and timely-throughput. The authors in \cite{huang2024aoiware} proposed a learning-based robust resource allocation considering overlapping interference and AoI-sensitive service requirements for ultra-dense Industrial IoT networks. Moreover, the authors in \cite{liu2024anaoi} constructed a blockchain-based remote intelligent healthcare system aiming to minimize both AoI and energy consumption of medical data transmission. The authors in \cite{lin2024aoi} investigated AoI impact on task-oriented multicasting in multi-cell non-orthogonal multiple access (NOMA) networks. The authors in \cite{huang2024aoiguaranteed} tackled the AoI-guaranteed transmission scheduling problem as an AoI-guaranteed multi-armed bandit problem. The authors in \cite{li2024aoi} proposed an AoI-aware waveform design scheme for cooperative joint radar-communications systems. The authors in \cite{qi2024deep} proposed a reconfigurable intelligent surface-assisted Internet of vehicles network, minimizing AoI of vehicle-to-infrastructure links while prioritizing vehicle-to-vehicle payload transmission. The authors in \cite{zhang2024minimizing} studied high-speed railway mobile networks to optimize sensor scheduling and transmit power, thereby minimizing average AoI. The authors in \cite{shi2024enhancing} pioneered AoI use in autonomous driving systems, showing how optimizing AoI simultaneously minimizes response time and maximizes throughput. In addition, the authors in \cite{xie2024distributed} presented an analytical framework establishing a dual-action guideline for minimizing average AoI in random access networks. The authors in \cite{qin2025velocity} analyzed the impact of user velocity on peak AoI distribution for ground and aerial users. The authors in \cite{Akbari2024AoI} investigated a decentralized UAV-aided MEC system tailored for smart agricultural applications, where processing nodes utilize network function virtualization technology. They formulated a sophisticated methodology for efficient network function virtualization orchestration that concurrently minimizes critical performance metrics. The authors in \cite{Liu2022AoI} analyzed a UAV-assisted IoT ecosystem featuring multiple UAVs operating in a comprehensive cycle, i.e., launching from a central data center, gathering data from distributed ground SNs, distributing information to various users, and subsequently returning to the data center. However, in the aforementioned studies, the optimal strategy for AoI optimization varied in different application scenarios, and generic approaches often failed to address the unique characteristics and constraints of specific deployments.

\par Different from these existing works, this work uniquely leverages laser charging technology to fundamentally address the energy limitations that hamper extended UAV operations, enabling sustained information freshness over longer mission durations. Based on this, we propose an optimization framework specifically tailored for laser-powered scenarios, effectively balancing AoI minimization with energy efficiency considerations in the context of LBD-powered UAV systems.

\subsection{Optimization Methods for Various UAV-assisted IoT Networks}

\subsubsection{Static Optimization Methods}

\par Many existing works have applied static optimization techniques to UAV-assisted IoT networks. For example, the authors in \cite{shen2023convex} introduced a novel trajectory planning framework for quadrotors landing on aerial vehicle carriers, where they combined a quadrotor trajectory planning method based on lossless convexification theory with a sequential convex programming approach, enabling autonomous landing on both stationary and moving aerial vehicle carriers in 3D space. The authors in \cite{dai2025uav} employed the A* algorithm, a well-established path planning approach, as a component of their two-step optimization methodology for mobile nest path planning. Moreover, the authors in \cite{meng2024evolutionary} tackled the multi-UAV cooperative path planning challenge by formulating it as a constrained optimization problem and introducing the evolutionary state estimation-based multi-strategy jellyfish search algorithm to identify high-quality trajectories. The authors in \cite{peng2022constrained} developed a constrained decomposition-based multi-objective evolution algorithm to address the formulated constrained multi-objective optimization problem, which incorporated dual objective functions focused on energy-efficient offloading and safe path planning for UAVs. In addition, the authors in \cite{sun2024multi} proposed a joint task offloading, computation resource allocation, and UAV trajectory control (JTORATC) approach, where the task offloading sub-problem was resolved using distributed splitting and threshold rounding methods, the computation resource allocation sub-problem was addressed through Karush-Kuhn-Tucker optimization, and the UAV trajectory control sub-problem was solved via successive convex approximation (SCA) techniques. The authors in \cite{Zhou2024Optimizing} developed two sophisticated algorithms, variable particle swarm optimization and twin variable neighborhood particle swarm optimization, to jointly optimize power, bandwidth, and UAV trajectories with the objective of minimizing AoI. The authors in \cite{Gao2023AoI} established a comprehensive start-to-end strategy incorporating association and planning mechanisms to minimize AoIs of two SNs through a methodical iterative three-step process. Initially, they determine the locations of data collection points (CPs) at which the UAVs hover to collect data and establish the SN-CP association using a density-based clustering algorithm. Subsequently, they cluster the CPs to form CP clusters and establish the CP-UAV association. Finally, leveraging the outcomes from the previous steps, they optimize the flight trajectories of the UAVs through an improved ant colony optimization algorithm while accounting for limited endurance capability constraints. However, these static optimization methods struggled with the dynamic nature of UAV-IoT systems, computational complexity in large-scale scenarios, and difficulty in adapting to changing environmental conditions without complete system information.

\subsubsection{DRL-based Optimization Methods}

\par Many existing works have applied DRL techniques to optimize energy efficiency and flight trajectories in UAV-assisted IoT networks. For instance, the authors in \cite{zhu2022aerial} utilized a dual UAV system (CUAVs and MUAVs) and employed DRL to minimize mission completion time by optimizing charging schedules and travel paths. Moreover, the authors in \cite{yi2020deep} proposed a DRL-based algorithm for UAV-assisted data collection to determine the optimal flight trajectory of the UAV and transmission scheduling of ground IoTs. The authors in \cite{miao2024utility} developed a UAV-aided video transmission system based on MEC and implemented a DDPG algorithm to achieve continuous action control through policy iteration. The authors in \cite{ghomri2024drl} proposed a PPO agent to enhance energy efficiency while addressing far-near fairness in NOMA-UAV networks by simultaneously controlling UAV trajectory, transmit power, node association, and power allocation. The authors in \cite{dai2025uav} devised a two-step optimization method employing modified multi-step dueling double DQN and A* algorithms to minimize inspection time while maximizing energy efficiency. In addition, the authors in \cite{ullah2024sum} examined a wireless-powered communication network where a resource-constrained secondary node harvests energy from ambient RF signals, implementing DRL to jointly optimize EH time and transmit power. The authors in \cite{pan2025multi} investigated a UAV-assisted IoT network in which the UAV sequentially accesses IoTs, proposing a DRL algorithm for multi-objective collaborative optimization that generates optimal strategies based on device priorities and assigned weights. The authors in \cite{zhang2024generative} developed generative artificial intelligence agents for model formulation and subsequently implemented a mixture of experts (MoE) approach to design transmission strategies. Specifically, they harnessed large language models to construct an interactive modeling paradigm and employed retrieval-augmented generation to extract satellite expert knowledge that underpins mathematical modeling. Subsequently, through the integration of expertise from multiple specialized components, they introduced an MoE-PPO approach to address the formulated problem. The authors in \cite{Zhang2023Energy} investigated reconfigurable intelligent surface (RIS)-assisted simultaneous wireless information and power transfer networks with rate splitting multiple access. To address the non-convex problem comprising both discrete and continuous variables, they proposed a DRL-based approach utilizing the PPO framework. Unlike traditional optimization approaches that optimize beamforming vectors and phase shifts separately and alternatively, their proposed PPO-based approach optimizes all variables simultaneously in unison. The authors in \cite{Panahi2023Reinforcement} formulated a double Q-learning-based trajectory design framework that enables energy-constrained surveillance UAVs to determine the optimal sequence of firefighting UAVs to visit (optimal flying trajectory), thereby maximizing the number of informed firefighting UAVs while accounting for limited execution time constraints. However, these DRL approaches were predominantly single-agent, centralized solutions that were inherently unsuitable for distributed environments, limiting their scalability and adaptability in multi-UAV scenarios.

\par Moreover, many existing works have investigated MADRL frameworks to optimize coordination and resource allocation in complex UAV networks. For example, the authors in \cite{oubbati2022multiagent} examined the use of UAV-enabled flying energy sources, powered by ground laser chargers, to support a set of MUAVs through RF wireless charging. They determined the positions of these flying energy sources using a MADRL approach called the multi-agent deep deterministic policy gradient (MADDPG) method, balancing fairness and energy consumption optimization. Moreover, the authors in \cite{he2024multi} proposed a cloud-based task processing framework with MADRL that generates near real-time task offloading decisions based on partially knowable future information. The authors in \cite{wang2024uav} developed a double-level DRL approach within a divide-and-conquer framework, where upper-level DRL manages task allocation while lower-level DRL handles route planning. The authors in \cite{qin2024drl} introduced a multi-agent proximal policy optimization (MAPPO) algorithm for UAV collaborative caching and a matching-DRL solution for trajectory planning, power allocation, and channel reusing. The authors in \cite{yin2024joint} formulated a MADRL-based joint optimization scheme for trajectory, power control, user association, and subcarrier allocation in multi-UAV networks. The authors in \cite{li2024drl} implemented MADDPG to optimize task scheduling and UAV trajectory in UAV-based MEC. In addition, the authors in \cite{tarekegn2024centralized} devised a centralized MADRL algorithm to adjust aerial base station trajectories based on link quality estimations. The authors in \cite{Wei2024Minimizing} introduced a multi-agent cooperative DQN algorithm with delayed reward to minimize the AoI in scenarios where only one UAV can be charged at a time. The authors in \cite{Messaoudi2024UGV} implemented a MADRL method to optimally control the trajectories of both unmanned ground vehicles (UGVs) and UAVs, thereby jointly reducing their energy consumption, decreasing the AoI of IoTs, and ensuring timely charging of UAVs while preventing their failures. The authors in \cite{Betalo2024Multi} utilized a multi-agent deep Q-network (MADQN) algorithm to jointly optimize UAV trajectories, EH, task scheduling, and data offloading with the objective of minimizing the AoI and enhancing energy efficiency. Through the MADQN algorithm, UAVs can identify optimal data collection and EH decisions to minimize their energy consumption and efficiently gather data from multiple SNs, resulting in reduced AoI and improved energy efficiency. However, while these multi-agent approaches achieved remarkable success in UAV coordination, they were not specifically tailored for the unique challenges presented by LBD-powered systems, limiting their effectiveness in such specialized environments.

\par Different from these existing works, this work designs a novel LBD-powered UAV data collection system and proposes a corresponding MADRL framework specifically tailored for energy-efficient information gathering in IoT networks. The considered approach addresses the limitations of static optimization methods through adaptive learning-based techniques, overcomes the scalability constraints of single-agent DRL approaches through multi-agent coordination, and specifically optimizes for the unique characteristics of laser-powered environments.


\section{System Model}
\label{sec: system model}

\par In this section, we first introduce the overall architecture of the LBD-powered multi-UAV data collection system in IoT networks. Then, we detail the data transmission model from IoTs to UAVs, the laser charging model from LBDs to UAVs, the AoI model, and the propulsion energy consumption model of UAVs. Finally, we formulate the problem. Note that the main notations used in this paper are summarized in Table~\ref{tab:notation}.

\begin{table*}[t]
\caption{{{}\\ \textsc{Main Notations.}}}
\label{tab:notation}
\centering
\renewcommand{\arraystretch}{1.3}  
\resizebox{\textwidth}{!}{
\begin{tabular}{@{}ll|ll@{}}
\toprule
\multicolumn{2}{c|}{\textbf{Notation used in system model}} & \multicolumn{2}{c}{\textbf{Notation used in reinforcement learning}} \\ \midrule
\textbf{Notation} & \textbf{Definition} & \textbf{Notation} & \textbf{Definition} \\ \midrule
$A$ & Rotor disc area & $\mathcal{A}$ & Action space \\
$d_0$ & Fuselage drag ratio & $\mathcal{A}_t$ & Action space for all UAVs at time slot $t$ \\
$E_{i}$ & Initial energy of each IoT & $a_t^i$ & Action of UAV $i$ at time slot $t$ \\
$E_{u}$ & Initial energy of each UAV & $f_{actor}$ & Policy head that outputs action probabilities \\
$H$ & Flight altitude of UAVs & $h_t^i$ & Hidden state that captures temporal dependencies \\
$\textit{N}_{\textit{D}}, \textit{N}_{\textit{S}}, \textit{N}_{\textit{U}}$ & The number of LBDs, IoTs and UAVs & $\mathcal{O}$ & Observation space \\
$P_{i}$ & Transmit power of the $i$-th IoT & $\mathcal{O}_t$ & Observation space for all UAVs at time slot $t$\\
$P_{L}$ & Laser transmitting power & $o_t^i$ & Local observation of UAV $i$ at time slot $t$ \\
$P_{\alpha}, P_{\beta}$ & Constants representing the blade profile power and induced power during hover & $\mathcal{R}$ & Immediate reward \\
$R_{c}$ & Radius of charging area & $r_t$ & Comprehensive reward function at time slot $t$ \\
$t, T, \mathcal{T}$ & The index, the number, and the set of time slots & $\mathcal{S}$ & State space \\
$t_d$ & Time duration of time slot & $\mathcal{S}_t$ & System state at time slot $t$ \\
$v$ & Flight speed of UAVs & $V_{local}^i$ & Individual performance value of UAV $i$ \\
$v_0$ & Mean rotor induced velocity in hover & $V_{global}$ & Overall system state value \\
$v_{\textsl{tip}}$ & Tip speed of the rotor blade & $\mathcal{O}$ & Observation space \\
$V_{i}$ & The amount of data to be transmitted each time generated by each IoT & $\mathcal{O}_t$ & Observation space for all UAVs at time slot $t$\\
$W$ & Channel bandwidth & $\theta$ & Parameter of LSTM-based actor networks for decentralized execution\\
$x_{i}^{s}, y_{i}^{s}, z_{i}^{s}$ & Coordinate of the $i$-th IoT & $\phi$ & Parameter of centralized critic network for global state evaluation\\
$x_{j}(t), y_{j}(t), z_{j}(t)$ & Coordinate of the $j$-th UAV in time slot $t$ & $\theta_{old}$ & Parameter of old actor network for stable policy updates\\
$x_{k}^{d}, y_{k}^{d}, z_{k}^{d}$ & Coordinate of the $k$-th LBD & $\gamma$ & discount factor \\
$\beta _{0}$ & Channel power at the reference distance 1 meter & $\pi_\theta$ & Police network \\
$\delta$ & Laser attenuation coefficient & $\omega_l, \omega_g$ & Learnable weights that balance local and global objectives \\
$\eta _{le}$ & Laser-to-electricity conversion efficiency & & \\
$\rho$ & Air density & & \\
$\sigma ^{2}$ & Gaussian noise power at the UAVs & & \\
$\omega$ & Rotor solidity & & \\
\bottomrule
\end{tabular}
}
\end{table*}

\subsection{System Overview}

\par
\begin{figure}[t]
\centerline{\includegraphics[scale=0.36]{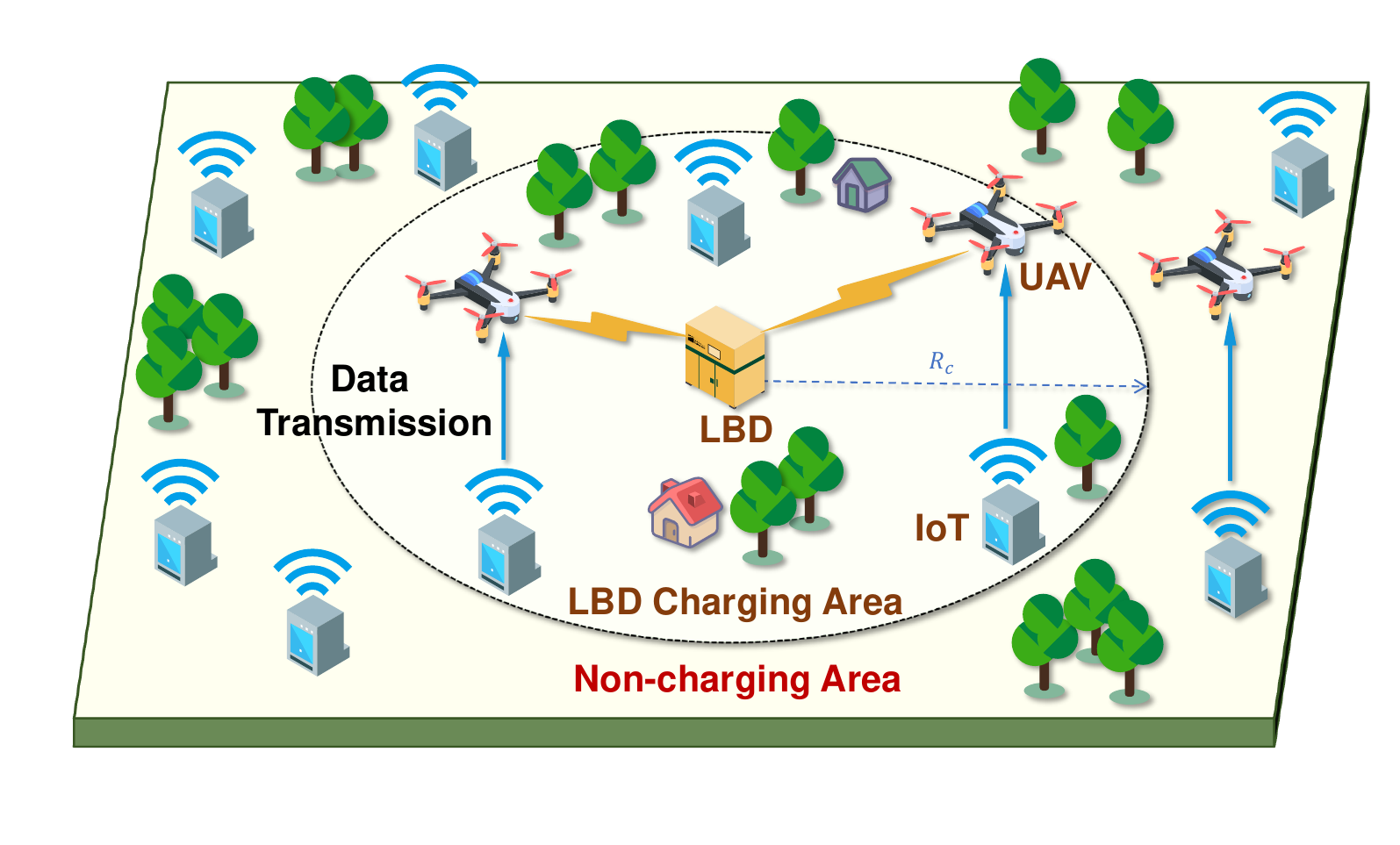}}
\caption{Sketch map of the LBD-powered multi-UAV data collection system in IoT networks.}
\label{fig1}
\end{figure}

 As illustrated in Fig.~\ref{fig1}, we consider an LBD-powered multi-UAV data collection system in IoT networks. Specifically, the system comprises two primary components, which are a set of IoTs denoted as $\mathcal{S} = \{1, 2, ..., \textit{N}_{\textit{S}}\}$ and a set of UAVs denoted as $\mathcal{U} = \{1, 2, ..., \textit{N}_{\textit{U}}\}$. Each IoT $s_{i} \in S$ is placed at a known fixed location. We also consider that each IoT stores $V_{i}$ units of data, which are static and do not increase over time, and has an initial energy level $E_{s}$. Moreover, the UAVs are tasked with visiting these IoTs to collect the stored data. We consider that each UAV $u_{j} \in U$ has a fixed initial energy $E_{u}$ and operates at a constant speed $v$ at a fixed altitude $H$. In addition, LBDs are denoted as $\mathcal{D} = \{1, 2, ..., \textit{N}_{\textit{D}}\}$. The recharge process occurs when the UAV is within the charging area defined by radius $R_{c}$ around an LBD, as shown in Fig.~\ref{fig1}. Consequently, UAVs are deployed to collect data while operating under energy limitations.

\par We consider a discrete-time system that evolves in time slots $\mathcal{T} \triangleq\{1,2, \ldots, T\}$, where the length of a time slot is equal to $t_d$ seconds. In this system, the UAVs collect data from any IoT when within its communication range. Since the energy of a UAV is depleted during flight and data communication, to prevent UAVs from exhausting their energy, they can be recharged at designated LBDs located within the operational area. Furthermore, data collection is considered to be instantaneous when the UAV is in the vicinity of an IoT.

\par We define a Cartesian coordinate system to describe the locations of IoTs, UAVs, and LBDs. Specifically, the IoTs are fixed at $(x_{i}^{s}, y_{i}^{s}, 0)$, representing their position at ground level. Moreover, UAVs fly at a constant altitude $H$, and their coordinates are represented as $(x_{j}(t), y_{j}(t), H)$ at any given time $t$, with $(x_{j}(t), y_{j}(t))$ denoting the horizontal position. Furthermore, the LBDs are located at known fixed coordinates $(x_{l}^{d}, y_{l}^{d}, z_{l}^{d})$, and the charging area is defined by a radius $R_{c}$ centered on each LBD, as illustrated in Fig.~\ref{fig1}. Therefore, the operational area includes regions designated for IoT communication, UAV operation, and charging. 

\par To establish a tractable theoretical framework for this complex system while ensuring meaningful optimization in the system, we make the following key assumptions:

\begin{itemize}

    \item \textit{The UAVs operate at a fixed altitude $H$ with constant speed $v$, enabling predictable flight patterns and simplifying trajectory optimization.} This assumption is commonly adopted in UAV trajectory optimization studies~\cite{Li2020Joint, Wu2022Route, Wu2020Optimal, li2025securing} as it reduces the complexity of the 3D path planning problem while remaining practical for many real-world UAV applications.
    
    \item \textit{All IoTs are stationary with fixed locations on the ground plane, each storing a static amount of data Vi that could be collected by UAVs.} Note that this is reasonable for most IoT monitoring scenarios such as environmental monitoring, smart agriculture, and infrastructure inspection, where sensors are deployed at strategic locations to collect environmental or infrastructure data~\cite{Zhang2022Joint, Wang2020Energy, Lu2024CODE, yuan2025ground}.
    
    \item \textit{Data collection is instantaneous when a UAV is within communication range of an IoT, allowing for efficient modeling of the collection process.} This assumption is widely adopted in UAV-assisted IoT networks and is reasonable given the high mobility of UAVs and the relatively small data packet sizes typical in IoT applications~\cite{Wei2022UAV, Xu2021Blockchain, Lu2023Federated, duan2025moco}.
    
    \item \textit{LBDs are deployed at fixed locations, creating circular charging areas with radius $R_c$ where UAVs can be recharged while continuing their operations.} This assumption follows established laser charging system models~\cite{Zhang2018Distributed, Jaafar2021Dynamics} and reflects the practical deployment of ground-based laser charging infrastructure.
    
    \item \textit{Laser charging from LBDs to UAVs is stable and reliable within the charging area, with charging power following the laser attenuation model without interruptions or fluctuations.} This assumption allows us to establish baseline performance and is supported by recent advancements in laser charging technology that have demonstrated high reliability in controlled environments~\cite{Wei2022UAV, Zhao2020Efficiency}.
    
\end{itemize}

\subsection{Data Transmission Model}

\par Let $P_{i}$ denote the transmit power of $s_{i}\in S$. Then, the data transmission rate from the IoT $s_{i}$ to UAVs in bits/Hz at time slot $t\in\mathcal{T}$ is given by
\begin{equation}
\label{eq:R_ij}
\begin{aligned}
R_{ij}^{f}(t)=W\log_{2}(1+\frac{\beta_{0}P_{i} [P^{\mathrm{LoS}}_{i,j}(t) \mu_{\mathrm{LoS}}+P^{\mathrm{NLoS}}_{i,j}(t) \mu_{\mathrm{NLoS}}] }{\sigma^{2}{d_{ij}(t)}^{\alpha}}),
\end{aligned}
\end{equation}

\noindent where $d_{ij}(t)=\sqrt{(x_{j}(t)-x_{i}^{s})^2+(y_{j}(t)-y_{i}^{s})^2}$ represents the distance between UAV $j$ and IoT $i$ at time $t$. Moreover, $W$ represents the channel bandwidth, $\beta _{0}$ denotes the channel power at the reference distance of 1 meter, and $\sigma ^{2}$ is the Gaussian noise power in UAVs. In addition, $P^{\mathrm{LoS}}_{i,j}(t)$ and $P^{\mathrm{NLoS}}_{i,j}(t)$ are the probability of LoS connection between sensor $i$ and UAV $j$. Specifically, $P^{\mathrm{LoS}}_{i,j}(t)=\left(1+b_1 e^{-b_2(\theta_{i, j}(t)-b_1)} \right)^{-1}$, and $P^{\mathrm{NLoS}}_{i,j}(t)=1-P^{\mathrm{LoS}}_{i,j}(t)$, in which $b_{1}$ and $b_{2}$ represent environment-dependent constants, and $\theta_{i, j}(t)$ is the angel between the IoT $i$ and the UAV $j$. 

\subsection{Laser Charging Model} 

In the charging area, each UAV can be charged with one LBD. Let the radius of the charging area on a horizontal plane be $R_{c}$. Then, the power received from the UAV $u_{j}\in U$ from any LBD at time slot $t\in\mathcal{T}$ is given by

\begin{equation}
P_{j}^{f}\left ( t \right )=P_{L}\cdot\eta _{le}\cdot e^{-\delta \cdot \sqrt{(x_{j}\left ( t \right )-x^{b})^{2}+(y_{j}(t)-y^{b})^{2}+H^{2}}},  
\label{eq:laser}
\end{equation}

\noindent where $\delta$ is the laser attenuation coefficient and $\eta_{le}$ denotes the laser-to-electricity conversion efficiency. Moreover, $P_{L}$ is the power of the laser.

\subsection{AoI Model}

The AoI depicts the time difference between the time an IoT generates data to be transmitted and the time data collection begins for this IoT. At first, all IoTs are equipped with information and ready to be accessed by UAVs. The initial value of the instant of time is equal to 0. In time slot $t$, when a UAV accesses an IoT that has information to transmit, the AoI of IoT $a_(t)=t-t'$, while $t'$ represents the time when the IoT last generated information. Let $a _{i}(t)$ be the AoI of the IoT $s_{i}$ at time instant $t$. When an IoT is accessed by a UAV and generates a larger AoI, the AoI of this IoT should be updated. Thus, the peak AoI of the network is given by
\begin{equation}
A = \max_{Q(t), T} \{a_1(t), a_2(t), \dots, a_n(t)\}.
\end{equation}

\subsection{Propulsion Energy Consumption Model of UAVs}

\par According to dynamic principles, the propulsion power consumption of a UAV can be modeled as a function of its velocity. In this study, we adopt the propulsion power consumption model presented in \cite{zeng2019energy}, which is given by

\begin{align}
P(v) & =  P_{\alpha} \left( 1 + \frac{3v^2}{v_{\textsl{tip}}^2} \right) \nonumber  + P_{\beta} \left( \sqrt{ 1 + \frac{v^4}{4v_0^4} - \frac{v^2}{2v_0^2} } \right)^{\frac{1}{2}} \nonumber \\  & + \frac{1}{2} d_0 \rho \omega A v^3,
\label{eq:energy consumption}
\end{align} 

\noindent where \( P_{\alpha} \) and \( P_{\beta} \) are constants, with \( P_{\alpha} \) representing the blade profile power and \( P_{\beta} \) representing the induced power during hover. Additionally, \( v_{\text{tip}} \) denotes the tip speed of the rotor blade, while \( v_0 \) is the mean rotor induced velocity in hover. The parameter \( d_0 \) corresponds to the fuselage drag ratio, and \( \omega \) represents the rotor solidity. Moreover, \( \rho \) denotes the air density, and \( A \) is the area of the rotor disc.

From Eq.~\eqref{eq:energy consumption}, it follows that the power consumption of the UAV at hover is $P(0) = P_{\alpha} + P_{\beta}$. The value of $P(0)$ is a constant that depends on the weight of the UAV, the air density, the rotor radius, and other factors. Based on Proposition 1 in \cite{wu2020load}, we conclude that Eq.~\eqref{eq:energy consumption} is convex for $v > 0$. Therefore, we can determine the optimal velocity $v_e = \arg \min P(v)$ that minimizes power consumption.

\subsection{Problem Formulation}

\par In this work, the primary optimization objective is to minimize the peak AoI across the LBD-powered multi-UAV data collection system in IoT networks. To achieve this goal, we need to carefully plan the flight trajectories of UAVs so that they can collect IoT information as efficiently and quickly as possible while adhering to charging strategies, thereby minimizing the peak AoI of the system.

\par To achieve the aforementioned optimization objective, we define the following key decision variables, including $\boldsymbol{A}=\left\{\left[a_u^x\left(t\right), a_u^y\left(t\right)\right] \mid t \in \mathcal{T}, u \in \mathcal{U}\right\}$ and $\boldsymbol{C}=\left\{c_u\left(t\right) \mid t \in \mathcal{T}, u \in \mathcal{U}\right\}$. Specifically, $\boldsymbol{A}$ is a matrix representing the control parameters of the UAVs, which denotes its spatial displacement at different time slots. On the other hand, $\boldsymbol{C}$ is a matrix representing the charging parameters of the UAVs, indicating whether the UAVs should leave or fly to the charging area, or continue to perform the missions at different time slots.

Following this, the optimization problem is formulated as follows:

\begin{subequations}
\begin{align}
    & \min_{A,C} && A = \max \{a_1(t), a_2(t), \dots, a_n(t)\}, \\
    & \ \ \ \text{s.t.} && \int_{0}^{T}R_{ij}^{f}(t)dt\geq V_{i}, t\in \mathcal{T} , 1\leq i\leq n, 1\leq j\leq m, \label{eq:b} \\
    & && E_{i}-\int_{0}^{T}P_{i}(t)dt\geq E_{\theta }, t\in \mathcal{T}, 1\leq i\leq n, \label{eq:c} \\
    & && 0< E_{j}(t)\leq E, t\in \mathcal{T}, 1\leq j\leq m, \label{eq:d} \\
    & && \sqrt{(x_{j}(t)-x_{{j}'}(t))^{2}+(y_{j}(t)-y_{{j}'}(t))^{2}}> d,\nonumber\\
    & && \ \ \ \ \ \ \ \ \ t\in \mathcal{T}, 1\leq j\leq m, 1\leq {j}'\leq m, j\neq {j}', \label{eq:e} \\
    & && \sqrt{(x_{j}\left ( t \right )-x^{b})^{2}+(y_{j}(t)-y^{b})^{2}}\leq 2R_{c},\nonumber\\
    & && \ \ \ \ \ \ \ \ \ \ \ \ \ \ \ \ \ \ \ \ \ \ \ \ \ \ \ \ \ \ \ \ t\in \mathcal{T}, 1\leq j\leq m, \label{eq:f} 
\end{align}
\end{subequations}
where constraint~\eqref{eq:b} ensures that all data from IoTs can be collected by UAVs. Moreover, constraint~\eqref{eq:c} guarantees that the remaining energy of $s_{i}$ is greater than or equal to $E_{\theta }$. In addition, constraint~\eqref{eq:d} ensures that the remaining energy of each UAV is within a reasonable range. Furthermore, constraint~\eqref{eq:e} is about collision control between the UAVs, and constraint~\eqref{eq:f} limits the flight area.

\par As can be seen, the formulated optimization problem is NP-hard. To establish the computational complexity of our problem, we demonstrate its relationship to the well-known traveling salesman problem (TSP), which is NP-hard. Consider a simplified version of our problem with a single UAV, where energy constraints and collision avoidance are temporarily ignored. In this case, the problem reduces to finding the shortest path that visits each IoT exactly once, precisely the definition of TSP. Given that TSP is NP-hard, and our problem extends it by incorporating additional complexities such as multiple UAVs, energy constraints, charging strategies, and peak AoI minimization, we can conclude that our problem is also NP-hard.

\section{MADRL-based Approach}
\label{sec: MADRL-based approach}

\par In this section, we propose an MADRL-based approach to solve our optimization problem. To this end, we first introduce the basics of DRL. Then, we show the motivations and rationales for using DRL and reformulate the problem as a partially observable Markov decision process (POMDP). Finally, we introduce the proposed MAPPO-TM algorithm with several improvements.

\subsection{Overview of the basics of DRL}

\par DRL represents a machine learning approach that integrates reinforcement learning with deep neural networks to address complex decision-making challenges in dynamic environments~\cite{Nguyen2020Deep}. Central to DRL is the Markov decision process (MDP), which offers a mathematical framework for modeling sequential decision-making under uncertainty~\cite{guo2022Real, Duan2024MOTO}. Within an MDP, an agent interacts with an environment through discrete time steps, thereby making decisions that aim to maximize cumulative rewards. From a mathematical perspective, an MDP comprises a tuple $(S, A, P, R, \gamma)$, where $S$ denotes the state space, $A$ represents the action space, $P$ indicates the state transition probability function, $R$ signifies the reward function, and $\gamma$ is the discount factor. In this framework, the DRL agent aims to discover an optimal policy $\pi^*$ that maximizes the expected cumulative reward over time through environment interaction and policy refinement based on received rewards~\cite{Deng2022AUCTION}.

\subsection{Motivations and Rationales for Employing DRL}

\par This work focuses on minimizing the peak AoI in an LBD-powered multi-UAV data collection system in IoT networks by optimizing UAV flight trajectories and charging strategies. The formulated problem exhibits the following key attributes:

\begin{itemize}

\item \textit{Real-time Decision-making:} In the formulated problem, UAVs need to continuously adjust their flight paths and charging strategies in response to fluctuating AoI of IoTs, varying energy levels, and positions of neighboring UAVs, introducing considerable uncertainty into the decision-making process.

\item \textit{Long-term Optimization Objectives:} In the formulated problem, decisions made by UAVs at any given time step influence subsequent AoI and energy levels, requiring an approach that prioritizes long-term optimization goals rather than merely immediate benefits. 

\item \textit{Computational Complexity:} As demonstrated in Section~\ref{sec: system model}, the challenge is the NP-hard nature, which extends beyond the traveling salesman problem with added complexities including multiple UAVs, energy constraints, charging strategies, and peak AoI minimization. 

\end{itemize}

\par Conventional optimization methods struggle to deal with the above challenges of the formulated optimization problem. Firstly, the problem is with NP-hard nature and complex non-linear relationships between objectives and decision variables, which may render conventional optimization techniques ineffective (\textit{e.g.}, exhaustive approach or convex optimization~\cite{Nievergelt2000Exhaustive, boyd2004convex}). Secondly, the formulated problem requires sequential decision-making in a dynamic environment and trade-off among optimization objectives, which leads to poor performance of conventional algorithms such as evolutionary algorithms that rely on accurate prior knowledge~\cite{Bliss2014evolutionary}. Finally, the vast solution space generated by trajectory planning and charging decisions of multiple UAVs, combined with intricate energy consumption models, makes developing an efficient online algorithm impractical~\cite{Nikolos2003Evolutionary}. 

\par In this case, DRL provides important advantages for these optimization challenges, particularly in the dynamics of the considered scenario. First, DRL learns from environmental interactions through trial and error, thereby enabling adaptation to changing conditions. This powerful adaptation capability of DRL makes it especially effective in the considered scenarios where IoT data generation patterns, UAV energy states, and AoI are in constant flux. Moreover, the capacity of DRL to optimize for future rewards enables it to effectively balance conflicting objectives (\textit{i.e.}, peak AoI minimization and energy limitations) by considering long-term outcomes rather than solely immediate gains. Consequently, the robust generalization capabilities of DRL and its proficiency in learning under uncertainty make it well-suited for the considered environment.

\subsection{Necessary Principles of MADRL}

\par MADRL enables multiple agents to learn optimal strategies through environment interaction and inter-agent communication, maximizing their long-term rewards without complete a priori knowledge. Different from the conventional DRL focuses on single-agent scenarios, MADRL extends this framework to multi-agent systems where decisions of agents mutually influence their rewards. Thus, POMDP is often adopted in multi-agent settings due to the limited observability of the agent.

\par A POMDP is formally defined as a tuple $(\mathcal{S}, \{ \mathcal{A}_i \}_{i=1}^N, \mathcal{P}, \{ \mathcal{R}_i \}_{i=1}^N, \{ \mathcal{O}_i \}_{i=1}^N)$, where $\mathcal{S}$ represents the state space, $\mathcal{A}_i$ denotes action space of agent $i$, $\mathcal{P}$ indicates the state transition probability function, $\mathcal{R}_i$ defines immediate reward function of agent $i$, and $\mathcal{O}_i$ describes observation space of agent $i$.

\par In MADRL, each agent $i$ aims to find an optimal policy $\pi_i^\ast$ that maximizes $\mathbb{E}_{\pi_i}(G_{i,t})$, where $\mathbb{E}_{\pi_i}(\cdot)$ represents the expected value under policy $\pi_i$. To determine $\pi_i^\ast$, a Q-value function $Q_i(s, a_1, a_2, \dots, a_N)$ is introduced, estimating expected discounted cumulative reward of agent $i$ when actions $(a_1, a_2, \dots, a_N)$ are executed at state $s$. In the following, we reformulate the optimization problem as a POMDP.

\subsection{POMDP Formulation}

\par Based on the MADRL principles mentioned above, we now formulate our optimization problem as a POMDP by defining its key components to address the UAV trajectory optimization and data collection challenges.

\begin{itemize}
    \item \textbf{State space} $\mathcal{S}$: To fully capture the system dynamics, the state space must include all essential information for decision-making. Thus, we design the state space to consist of UAV positions for trajectory tracking, UAV energy levels for charging strategy, and AoI of IoTs and information status for data freshness monitoring. At time slot $t$, the system state is defined as 
    \begin{align}
        \mathcal{S}_t=\{ & \left(x_j(t), y_j(t)\right), E_j(t), a_i(t), s_i(t) \mid \nonumber \\
        & i=1,2, \ldots, N_S, j=1,2, \ldots, N_U\},     
    \end{align}

    where $(x_{j}(t), y_{j}(t))$ represents the horizontal coordinate of UAV $u_{j}\in \mathcal{U}$.

    \item \textbf{Observation space} $\mathcal{O}$: In practical scenarios, each UAV has limited sensing capabilities and can only obtain local information. Different from the global state space, we design the observation space to reflect this partial observability. Each UAV observes its own position, energy level, and only the information status and AoI of accessible or nearby IoTs. Thus, the observation of UAV $u_{j}$ at time slot $t$ is defined as 
    \begin{align}
        \mathcal{O}_t^j=\{ & \left(x_j(t), y_j(t)\right), E_j(t), a_i(t), s_i(t) \mid \nonumber \\
        & i=1,2, \ldots, N_S^j\}.
    \end{align}
     As such, the observation space for all UAVs at time slot $t$ is denoted as 
     \begin{equation}
         \mathcal{O}_t=\left\{\mathcal{O}_t^j \mid j=1,2, \ldots, N_U\right\}.
     \end{equation}

    \item \textbf{Action space} $\mathcal{A}$: To ensure practical implementation and reduce computational complexity, the horizontal movements of UAVs are limited to eight standard directions (i.e., north, northeast, east, etc.). This direction-based action design not only corresponds to realistic UAV control strategies, but also effectively captures the temporal dynamics of UAV motion. Furthermore, by constraining the action space to directional outputs rather than absolute positional coordinates, the bounded nature of the directional space enhances algorithmic convergence properties~\cite{Zhang2021Three, Mei20223D, Bayerlein2021Multi, Li2020Path}. The action of UAV $u_{j}$ at time slot $t$ is defined as $\mathcal{A}_{t}^{j}=\left \{ a_{j}^x(t), a_{j}^y(t)\right \}$, and the action space for all UAVs is denoted as 
    \begin{equation}
        \mathcal{A}_t=\left\{\mathcal{A}_t^j \mid j=1,2, \ldots, N_U\right\}.
    \end{equation}

    \item \textbf{Immediate reward} $\mathcal{R}$: To align with our optimization objective, the reward function is carefully designed to balance multiple goals. The immediate reward function $r_{t}$ considers both the peak AoI of the network and a penalty term $r_{p}(t)$ defined as follows:
    
    \begin{equation}
    r_p(t)=\left\{\begin{array}{lr}
    -d_c \cdot r_{p e n 1}, & \text { if } E_j(t) \leq E_\phi, \\
    -d_c \cdot r_{p e n 2}, & \text { if } E_j(t)=E, \\
    r_0, & \text { if } E_\phi<E_j(t)<E,
    \end{array}\right.
    \end{equation}
    
    where $d_c$ represents the distance from UAVs to the charging area boundary, $E_{\phi }$ denotes the UAV energy threshold for immediate charging, and $E$ indicates the full UAV energy level. This penalty design encourages efficient energy management by penalizing both low energy states (risking operation interruption) and full energy states (indicating inefficient charging).

    \par Following this, the comprehensive reward function is then designed as follows:
    \begin{equation}
    r_{t}= \alpha \cdot r_{a}(t) + \beta \cdot r_{p}(t) + \gamma \cdot r_s(t),
    \end{equation}
    where $r_{a}(t)$ represents the reward associated with the AoI in the network at time slot $t$, $r_s(t)$ denotes the reward for successful information collection from IoTs at time slot $t$, $\alpha$, $\beta$, and $\gamma$ are weight factors for different parts of the reward function, respectively. As can be seen, this design directly relates to our optimization goal by penalizing high AoI while encouraging timely data collection and proper energy management.
\end{itemize}

\par Through this POMDP formulation, we can transform our optimization problem into a format suitable for MADRL algorithms. In the following, we introduce and analyze the standard MAPPO algorithm.

\begin{figure*}[t]
  \centering
  \includegraphics[width=0.9 \textwidth]{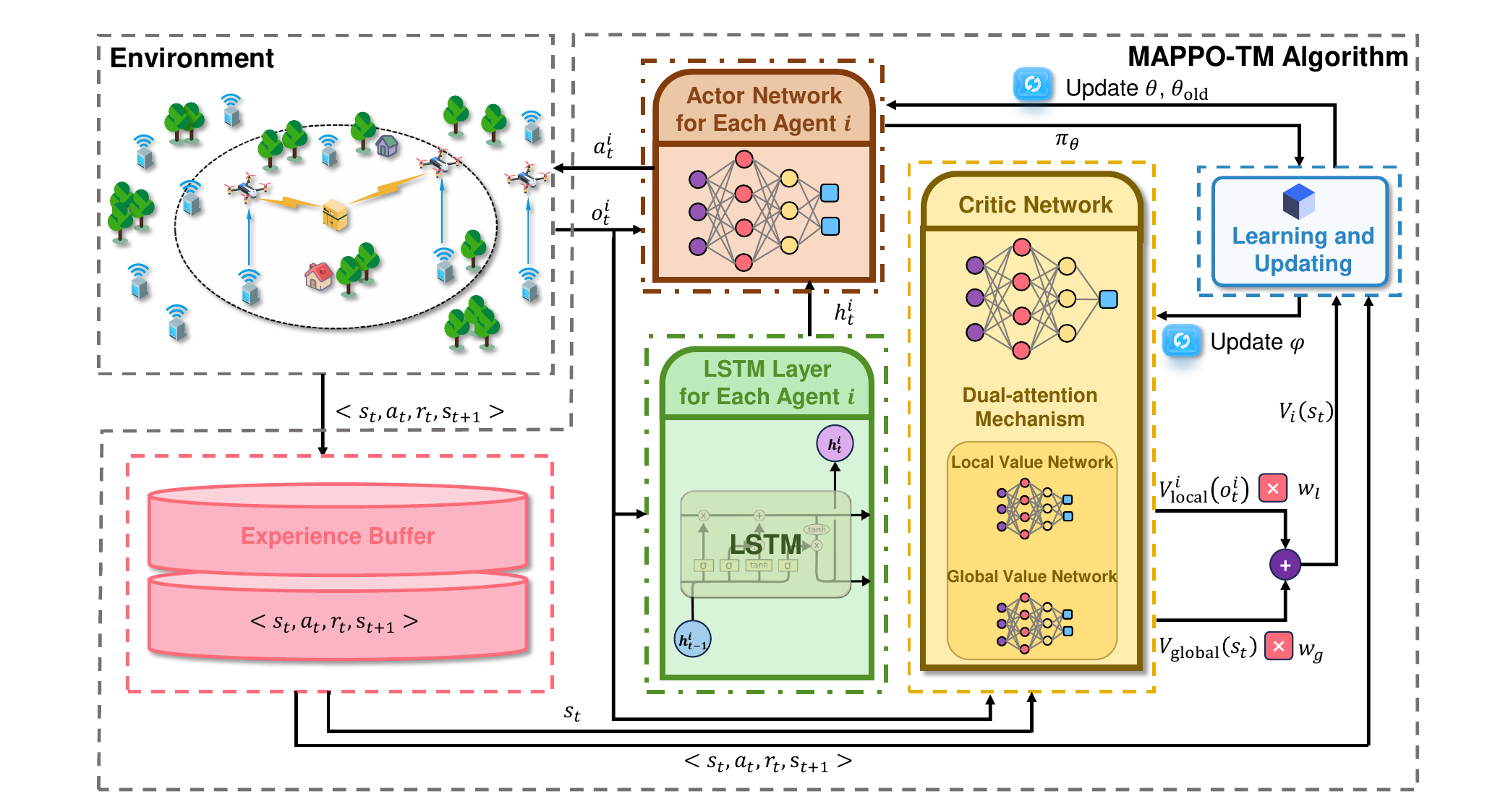}
  \caption{The framework of the proposed MAPPO-TM algorithm.}
  \label{fig:mappo_tm_overview}
\end{figure*}

\subsection{Standard MAPPO Algorithm}

\par In this work, we adopt MAPPO as our solution framework, which extends PPO to handle multi-agent scenarios effectively. The algorithm generates independent policy networks $\pi_{\theta_i}$ for each agent $i$, enabling decision-making based on their local observations. The optimization objective of MAPPO is given by
\begin{equation}
J(\theta) = \mathbb{E}_{\tau \sim \pi_{\theta}} \left[ \sum_{t=0}^{T} r_t \right],
\label{eq:MAPPO_objective}
\end{equation}
\noindent where $\tau$ denotes the joint trajectories under policy $\pi_{\theta}$, and $r_t$ represents the reward at time step $t$.

\par To address the scalability issue in MADRL, MAPPO employs centralized training with decentralized execution (CTDE). During training, a centralized critic evaluates the global state by minimizing the temporal difference error:
\begin{equation}
J_{\text{critic}}(V) = \mathbb{E}_{\tau} \left[ \sum_{t=0}^{T} \frac{1}{2} \left( r_t + \gamma V(s_{t+1}) - V(s_t) \right)^2 \right],
\label{eq:MAPPO_critic}
\end{equation}
\noindent where $\gamma$ is the discount factor and $V(s_t)$ represents the global value function at time step $t$.

\par Following the PPO design, MAPPO adopts a clipped surrogate objective for stable policy updates as follows:
\begin{equation}
L^{\text{clip}}_i(\theta) = \mathbb{E}_t \left[ \min \left( \rho_t^i(\theta) A_t^i, \text{clip}(\rho_t^i(\theta), 1-\epsilon, 1+\epsilon) A_t^i \right) \right],
\label{eq:MAPPO_clip}
\end{equation}
\noindent where $\rho_{t}^i(\theta)$ denotes the probability ratio, $A_t^i$ represents the advantage function for agent $i$, and $\epsilon$ controls the update step size.

\par However, the standard MAPPO algorithm faces two main challenges in our scenario. First, the specially designed reward function that balances AoI, energy management, and data collection may lead to unstable convergence during training. Second, the complex coordination among multiple UAVs to achieve efficient coverage while avoiding energy depletion requires more sophisticated policy updates. Thus, we propose several enhancements to the algorithm in the following section to address these challenges.

\subsection{MAPPO-TM Algorithm}

\par Based on the standard MAPPO, we propose MAPPO-TM to address the specific challenges in our UAV-enabled data collection scenario. 

\par The algorithm introduces two key enhancements, which are temporal dependency learning for capturing UAV historical trajectories and energy states, and multi-agent coordination mechanism for balancing individual and global objectives. These improvements effectively address the convergence and coordination issues faced by standard MAPPO in our POMDP formulation, and are detailed as follows.

\subsubsection{Temporal Dependency Learning}

\par In our POMDP formulation, the UAVs need to make decisions based on both current observations and historical information (e.g., previous trajectories and energy consumption patterns). However, standard MAPPO with feed-forward networks can only process current states, leading to suboptimal decisions in temporally dependent scenarios. To address this limitation, we incorporate long short-term memory (LSTM) networks into the actor network, so that allowing each UAV to maintain and utilize historical information effectively.

\par Specifically, the LSTM-enhanced actor network for agent $i$ is designed as follows:
\begin{equation}
h_t^i = \text{LSTM}(o_t^i, h_{t-1}^i; \theta_{\text{LSTM}}^i),
\end{equation}
\begin{equation}
\pi_{\theta_i}(a_t^i|o_t^i) = f_{\text{actor}}(h_t^i; \theta_{\text{actor}}^i),
\end{equation}
where $h_t^i$ represents the hidden state that captures temporal dependencies, $o_t^i$ is the current observation, and $f_{\text{actor}}$ denotes the policy head that outputs action probabilities.

\par By using LSTM-enhanced actor networks, we capture complex temporal dependencies across UAV trajectories, energy states, and data freshness metrics. This allows our agents to maintain critical historical context for decision optimization, unlike standard MAPPO which is limited to processing only current states through feed-forward networks. This enhancement enables each UAV to better plan its trajectory and energy usage by considering historical patterns.

\subsubsection{Multi-agent Coordination Mechanism}

\par Standard MAPPO faces challenges in balancing individual UAV objectives (energy management) with global goals (network AoI minimization) in our scenario. As depicted in Fig.~\ref{fig:mappo_tm_overview}, to enhance coordination, we introduce a dual-attention mechanism that allows each UAV to weigh both local and global information during decision-making. The coordination-enhanced value function is given by 
\begin{equation}
V_i(s_t) = w_l V_{\text{local}}^i(o_t^i) + w_g V_{\text{global}}(s_t),
\end{equation}
where $w_l$ and $w_g$ are learnable weights that balance local and global objectives, $V_{\text{local}}^i$ evaluates individual performance, and $V_{\text{global}}$ assesses the overall system state.

\par By implementing a dual-attention coordination mechanism, we enable each UAV to dynamically weigh local and global information during decision-making. This sophisticated coordination balances individual UAV objectives with system-wide performance metrics, particularly important when multiple UAVs must collectively optimize data freshness while managing individual energy constraints. This mechanism helps UAVs maintain energy efficiency while contributing to the global AoI optimization goal.

\par Through these enhancements, MAPPO-TM significantly improves upon standard MAPPO by enabling temporal-aware decision-making and better multi-agent coordination, leading to more efficient UAV trajectory planning and data collection strategies.

\subsubsection{Main Steps of MAPPO-TM Algorithm}

\par Following the POMDP formulation and algorithm design, we now present the main steps of MAPPO-TM, as shown in Algorithm \ref{alg:MAPPO-TM}. The algorithm begins by initializing the LSTM-based actor networks with parameters $\theta$ for decentralized execution, a centralized critic network with parameters $\phi$ for global state evaluation, and an old actor network with parameters $\theta_{\text{old}}$ for stable policy updates. For each episode, after environment reset and initial state $s_t$ acquisition, each UAV $i$ obtains its local observation $o_t^i$ and generates action $a_t^i$ through its LSTM-enhanced actor network, capturing temporal dependencies in the decision-making process. The environment then executes the joint action set $\mathbf{a}_t = \{a_t^1, a_t^2, \dots, a_t^N\}$, providing reward $r_t$ and next state $s_{t+1}$. These interaction experiences, including states, actions, rewards, and next states, are stored in an experience buffer for subsequent learning.

\par The learning process consists of two key components, which are critic network updates and actor network updates. The centralized critic network is updated by minimizing the value function loss, which evaluates global state values while considering both individual UAV performance and network-wide AoI optimization. Concurrently, the actor networks are updated using the clipped surrogate objective enhanced with our coordination mechanism, ensuring stable policy improvements while maintaining effective multi-agent cooperation. To further stabilize the training process, the old actor network parameters $\theta_{\text{old}}$ are periodically synchronized with the current parameters $\theta$, thereby maintaining consistent clipping ratios in the surrogate objective. This iterative process continues for $N_{\text{eps}}$ episodes, thus progressively improving both individual UAV policies and overall system performance.

\par By building upon the inherent stability with its clipped surrogate objective of PPO, we ensure controlled policy updates crucial for convergence in our complex environment. This stability foundation addresses the high dimensionality and stochastic elements of our problem domain, while our targeted enhancements overcome standard PPO limitations.

\subsection{Complexity Analysis}

\begin{table*}[t]
\caption{{{}\\ \textsc{Algorithms Complexity Comparison.}}}
\label{tab:complexity}
\centering
\renewcommand{\arraystretch}{1.3}  
\resizebox{\textwidth}{!}{
\begin{tabular}{@{}lll@{}}
\toprule
\footnotesize \textbf{Algorithms} & \textbf{Time Complexity} & \textbf{Space Complexity} \\
\midrule
MAPPO-TM \ \ \ \ \ \ \ \ \ \ \ \ \ \ \ & $\mathcal{O}(N_{\text{eps}}  T  N  |\theta_a| + N_{\text{eps}}  T  |\theta_c|)$ \ \ \ \ \ \ \ \ \ \ \ \ \ \ \ & $\mathcal{O}(N  |\theta_a| + |\theta_c| + D(|s| + |a| + 1))$ \\
MAPPO & $\mathcal{O}(N_{\text{eps}} T  N  |\theta_a'| + N_{\text{eps}}  T  |\theta_c'|)$ & $\mathcal{O}(N  |\theta_a'| + |\theta_c'| + D(|s| + |a| + 1))$ \\
MATD3 & $\mathcal{O}(N_{\text{eps}}  T  N  |\theta_a''| + N_{\text{eps}}  T  N  2  |\theta_c''|)$ & $\mathcal{O}(N  |\theta_a''| + N  2  |\theta_c''| + D(|s| + |a| + 1))$ \\
MADDPG & $\mathcal{O}(N_{\text{eps}}  T  N  |\theta_a'''| + N_{\text{eps}}  T  N  |\theta_c'''|)$ & $\mathcal{O}(2  N  |\theta_a'''| + 2  N  |\theta_c'''| + D(|s| + |a| + 1))$ \\

\bottomrule
\end{tabular}
}
\end{table*}

\begin{algorithm}[t]
\caption{MAPPO-TM}
\label{alg:MAPPO-TM}
Initialize the parameters $\theta$ of the LSTM-based actor networks and the parameters $\phi$ of the centralized critic network
\tcp*{Initialize temporal learning networks.}
Initialize the parameters of the old actor networks $\theta_{\text{old}} \leftarrow \theta$
\tcp*{For stable updates.}
\For{Episode $= 1, \dots, N_{\text{eps}}$}{
    Reset the environment and initialize the state $s_t$
    \tcp*{Start new episode.}
    \For{Time slot $t = 1, \dots, T$}{
        \For(\tcp*[f]{Decentralized execution phase.}){Each agent $i$}{           
            Obtain the current state $s_t^i$
            \tcp*{UAV local observation.}
            Pass $s_t^i$ through the LSTM-based actor to generate action $a_t^i$
            \tcp*{Temporal-aware decisions.}
        }
        Execute joint actions $\mathbf{a}_t = \{a_t^1, \dots, a_t^N\}$
        \tcp*{UAVs trajectory execution.}
        Receive reward $r_t$ and next state $s_{t+1}$
        \tcp*{AoI and energy-based reward.}
        Store transitions $(s_t, \mathbf{a}_t, r_t, s_{t+1})$ into the experience buffer\;
        Update the centralized critic using the value function loss
        \tcp*{With dual-attention mechanism.}
        Update the actor networks using the clipped surrogate objective\;
        Update $\theta_{\text{old}} \leftarrow \theta$ periodically
        \tcp*{Stabilize training.}
    }
}
\end{algorithm}

\par The computational complexity of the MAPPO-TM algorithm can be analyzed in both the training and execution phases. 

\begin{itemize}

    \item \textbf{\textit{For the training phase}}, the time complexity consists of several key components. First, the network initialization requires $\mathcal{O}(N |\theta_a| + |\theta_c|)$ computations for both actor networks and centralized critic network, where $|\theta_a|$ and $|\theta_c|$ represent their respective parameter sizes. Second, the action sampling process through LSTM-based actor networks takes $\mathcal{O}(N_{\text{eps}} T N |\theta_a|)$ operations, where $N_{\text{eps}}$ denotes the training episodes, $T$ represents time slots per episode, and $N$ is the number of UAVs. Third, the experience storage requires $\mathcal{O}(N_{\text{eps}} T N V)$ operations for maintaining the replay buffer, where $V$ denotes the size of state-action-reward tuples. Finally, the network update process needs $\mathcal{O}(N_{\text{eps}} T |\theta_c| + N_{\text{eps}} T N |\theta_a|)$ computations for both critic and actor networks. Thus, the overall time complexity during training is $\mathcal{O}(N_{\text{eps}} T N |\theta_a| + N_{\text{eps}} T |\theta_c|)$.

    \par The space complexity during training mainly comes from two aspects, which are network parameters storage and replay buffer maintenance. The former requires $\mathcal{O}(N |\theta_a| + |\theta_c|)$ space for storing all network parameters, while the latter needs $\mathcal{O}(D(|s| + |a| + 1))$ space for maintaining the replay buffer with size $D$, where $|s|$ and $|a|$ represent the dimensions of state and action spaces, respectively. Therefore, the total space complexity in training phase is $\mathcal{O}(N |\theta_a| + |\theta_c| + D(|s| + |a| + 1))$.

    \item \textbf{\textit{For the execution phase}}, since only the actor networks are utilized for action sampling, the time complexity reduces to $\mathcal{O}(T N |\theta_a|)$ and the space complexity becomes $\mathcal{O}(N |\theta_a|)$. This significant reduction in computational complexity makes MAPPO-TM suitable for real-world UAV deployments.
\end{itemize}

\par In addition, the complexity comparison between the proposed MAPPO-TM algorithm and other conventional MADRL algorithms is shown in Table~\ref{tab:complexity}. As can be seen, MAPPO-TM offers a balanced computational complexity compared to other MADRL methods. The efficiency-performance balance makes MAPPO-TM particularly well-suited for the UAV trajectory optimization and data collection challenges addressed in this work.

\section{Simulation Results}
\label{sec: simulation results}

\par In this section, we present simulation results and analyses. We first introduce the simulation setup and benchmarks, and then provide simulation results.

\subsection{Simulation Setups}

\begin{figure*}[t]
  \centering
  \begin{subfigure}[b]{0.32\textwidth}
    \includegraphics[width=\textwidth]{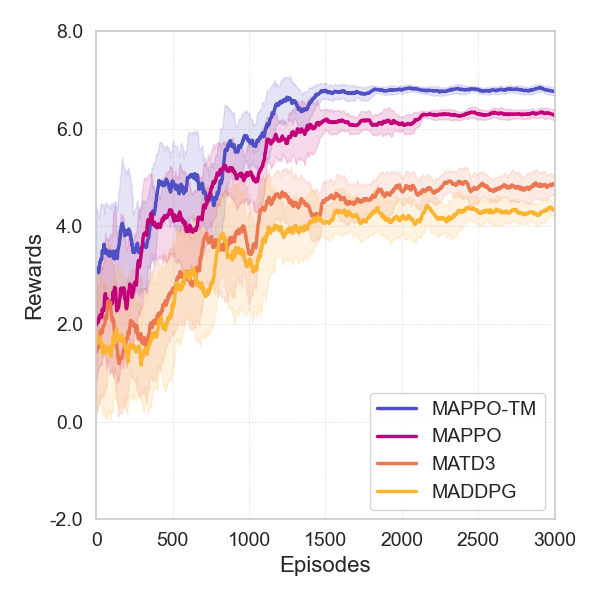}
    \caption{}
    \label{fig3:subfig1}
  \end{subfigure}
  \hfill
  \begin{subfigure}[b]{0.32\textwidth}
    \includegraphics[width=\textwidth]{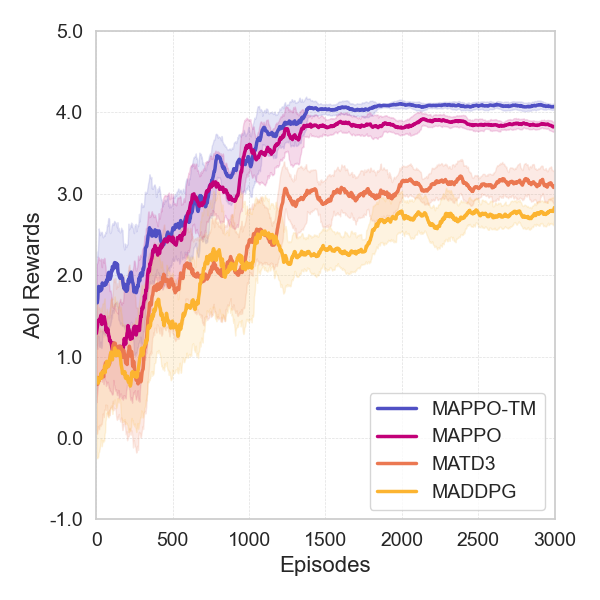}
    \caption{}
    \label{fig3:subfig2}
  \end{subfigure}
  \hfill
  \begin{subfigure}[b]{0.32\textwidth}
    \includegraphics[width=\textwidth]{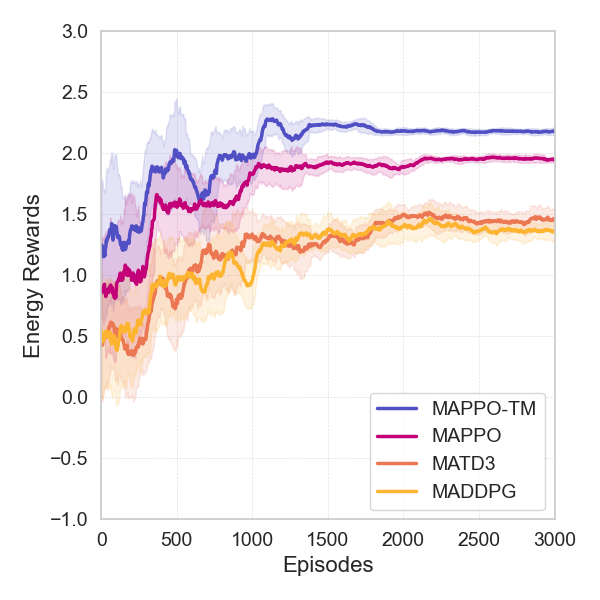}
    \caption{}
    \label{fig3:subfig3}
  \end{subfigure}
  \caption{Training results. (a) Cumulative rewards training curve. (b) AoI rewards training curve. (c) Energy rewards training curve.}
  \label{fig3}
\end{figure*}

\par We consider a typical field environment with multiple IoTs and LBDs. The UAVs fly in the airspace above the square area. The scene is divided into a circular charging area at the center and the rest of the non-charging area. The coordinate system takes the center of the square area, i.e., the coordinates of the LBD, as the origin. The initial coordinates of the IoTs are fixed, while the initial coordinates of the four UAVs are (1, 0, $H$), (-1, 0, $H$), (0, 1, $H$), and (0, -1, $H$), respectively. Furthermore, the initial power of the UAVs is 60\% of the full power. Other parameters are shown in Table~\ref{tab2}.

\par For comparison, we utilize MAPPO\cite{yu2022surprising}, multi-agent twin delayed deep deterministic policy gradient (MATD3)\cite{ackermann2019reducing}, and MADDPG \cite{lowe2017multi} as benchmark methods.

\begin{itemize}
    \item MAPPO: MAPPO extends the PPO framework to multi-agent environments. It employs CTDE, allowing each agent to learn its policy while leveraging global information during training. MAPPO maintains the stability and sample efficiency of PPO, making it suitable for complex multi-agent tasks with continuous action spaces. Additionally, it incorporates mechanisms to handle the non-stationarity arising from multiple learning agents, ensuring robust and scalable policy learning.

    \item MATD3: MATD3 adapts the twin delayed deep deterministic policy gradient (TD3) algorithm for multi-agent settings. It utilizes multiple critic networks to mitigate overestimation biases and employs a delayed update strategy to stabilize training across agents. MATD3 leverages decentralized policies while maintaining centralized critics, enabling effective coordination among agents in environments with high noise and uncertainty. This approach enhances the accuracy of value estimates and promotes cooperative behavior in multi-agent scenarios.

    \item MADDPG: MADDPG extends the DDPG algorithm to multi-agent systems. It employs CTDE, where each agent has its own actor network and its own centralized critic network that considers the observations and actions of all agents. This framework helps address the non-stationarity problem in multi-agent environments by providing each agent with additional information during training. MADDPG is effective in continuous action spaces and facilitates coordinated strategies among multiple agents, improving overall performance in complex tasks.
    
\end{itemize}

\begin{table}[tb]
\caption{{\\ \textsc{Simulation Parameters.}}}
\begin{center}
\renewcommand{\arraystretch}{1.6}  
\newcolumntype{P}[1]{>{\centering\arraybackslash}m{#1}} 
\newcolumntype{Q}[1]{>{\centering\arraybackslash}m{#1}} 
\begin{tabular}{|P{1.5cm}|Q{1.5cm}|P{1.5cm}|Q{1.5cm}|} 
\hline \textbf{Parameter} & \textbf{Value} & \textbf{Parameter} & \textbf{Value} \\
\hline $A$ & 0.1256 m\textsuperscript{2} & $R_{c}$ & 250 m\\
\hline $d_0$ & 0.5009 & $t_d$ & 1 s\\
\hline $E_{u}$ & 30000 J & $v$ & 5 m/s\\
\hline $H$ & 80 m & $v_0$ & 5.0463 m/s\\
\hline $\textit{N}_{\textit{D}}$ & 10 & $v_{\textsl{tip}}$ & 80 m/s\\
\hline $\textit{N}_{\textit{S}}$ & 50 & $\delta$ & 10\textsuperscript{-6}\\
\hline $\textit{N}_{\textit{U}}$ & 4 & $\eta _{le}$ & 0.15\\
\hline $P_{L}$ & 1000 W & $\rho$ & 1.225 kg/m\textsuperscript{3}\\
\hline $P_{\alpha}$ & 14.7517 W & $\frac{\beta_0}{\sigma^2}$ & 80 dB\\
\hline $P_{\beta}$ & 41.5409 W & $\omega$ & 0.1248\\
\hline
\end{tabular}
\label{tab2}
\end{center}
\end{table}

\subsection{Simulation Results}

\begin{figure*}[t]
  \centering
  \begin{subfigure}[b]{0.3\textwidth}
    \includegraphics[width=\textwidth]{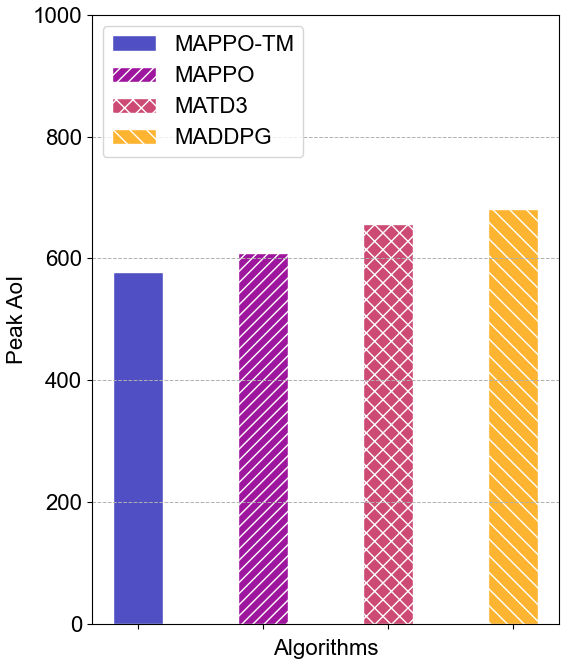}
    \caption{}
    \label{fig4:subfig1}
  \end{subfigure}
  \hfill
  \begin{subfigure}[b]{0.32\textwidth}
    \includegraphics[width=\textwidth]{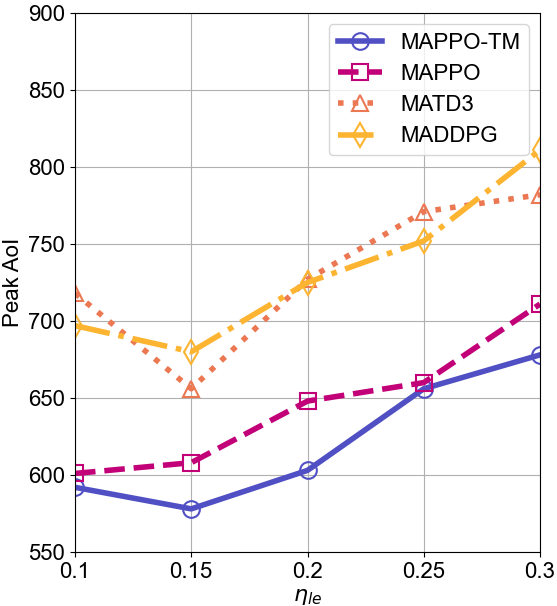}
    \caption{}
    \label{fig4:subfig2}
  \end{subfigure}
  \hfill
  \begin{subfigure}[b]{0.34\textwidth}
    \includegraphics[width=\textwidth]{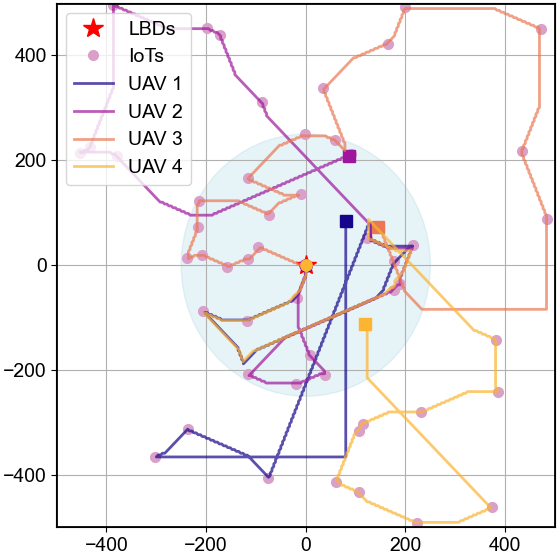}
    \caption{}
    \label{fig4:subfig3}
  \end{subfigure}
  \caption{(a) Peak AoI comparison across algorithms. (b) Peak AoI changing across different values of $\eta _{le}$. (c) Flight Paths of UAVs driven by MAPPO-TM}
  \label{fig4}
\end{figure*}

\par In this subsection, we evaluate the performance of the proposed MAPPO-TM framework under various UAV-enabled data collection scenarios. 

\subsubsection{Reward Performance Analysis}

\par Fig. \ref{fig3:subfig1} presents a comparative analysis of the four algorithms in terms of cumulative rewards. The results demonstrate that during the early training stages, the proposed MAPPO-TM algorithm maintains relatively stable performance at a higher reward level, while the conventional MAPPO algorithm performs marginally below MAPPO-TM. In contrast, MATD3 and MADDPG exhibit considerable fluctuations at lower reward levels, with MADDPG occasionally experiencing negative rewards. As training progresses, MAPPO-TM significantly outperforms the other algorithms. This superior performance can be attributed to the LSTM-based temporal dependency learning mechanism, which enables each UAV to effectively maintain and utilize historical information, thereby enhancing reward stability during initial training phases. Furthermore, the dual-attention coordination mechanism guides UAVs to balance local and global objectives, thus optimizing network-wide performance and maximizing cumulative rewards.

\par Fig.~\ref{fig3:subfig2} depicts the AoI rewards over 3000 episodes for MAPPO-TM, MAPPO, MATD3 and MADDPG. MAPPO-TM significantly outperforms the baselines, stabilizing at an AoI reward of approximately 4.1 after 1500 episodes, while MAPPO plateaus at around 3.9, MATD3 at 3.2, and MADDPG at 2.6. The shaded regions around each line indicate variance, with MAPPO-TM showing the smallest variance, reflecting stable performance, whereas MAPPO, MATD3, and MADDPG exhibit larger fluctuations, particularly in the early stages, with MADDPG displaying the highest variability throughout. The superior AoI performance of MAPPO-TM is primarily due to its multi-agent coordination mechanism, which uses a dual-attention approach to balance individual UAV objectives with the global goal of AoI minimization, thereby enabling more effective data collection strategies. Additionally, the LSTM-based temporal dependency learning allows MAPPO-TM to leverage historical trajectories, addressing the limitations of the feed-forward networks in standard MAPPO, which struggle with long-term planning, and the lack of coordination in MATD3 and MADDPG, thereby leading to their lower AoI rewards.

\par Fig.~\ref{fig3:subfig3} illustrates the energy rewards over 3000 episodes for the same four algorithms. MAPPO-TM again achieves the highest energy rewards, stabilizing at around 2.2 after 1800 episodes, while MAPPO plateaus at 2.0, MATD3 at 1.4, and MADDPG at 1.3. The variance, indicated by the shaded regions, is smallest for MAPPO-TM, suggesting consistent energy efficiency, whereas MAPPO shows moderate fluctuations, and MATD3 and MADDPG exhibit significant variability, with MADDPG performing the least reliably. The enhanced energy efficiency of MAPPO-TM can be attributed to its LSTM-enhanced actor network, which captures historical energy consumption patterns, allowing UAVs to make informed decisions that optimize energy usage over time, different from the baselines that lack temporal awareness. Furthermore, the coordination mechanism of MAPPO-TM ensures that UAVs balance energy management with global objectives, thus preventing the energy depletion issues seen in MATD3 and MADDPG, which struggle to coordinate effectively, and in MAPPO, which cannot account for long-term energy dependencies, thereby resulting in their lower energy rewards.

\subsubsection{Peak AoI Performance Evaluation}

\par To further validate the effectiveness of MAPPO-TM, we analyze peak AoI performance across the different algorithms. Fig.~\ref{fig4:subfig1} illustrates the final peak AoI values, with MAPPO-TM achieving significantly lower AoI compared to MAPPO and substantially outperforming other algorithms, while MATD3 and MADDPG demonstrate considerably higher peak AoI values respectively. This superior performance of MAPPO-TM can be attributed to two key factors. First, the LSTM network effectively captures temporal dependencies in UAV trajectories and energy states, enabling more informed decision-making and reducing the probability of suboptimal actions, consequently lowering peak AoI. Second, the dual-attention mechanism successfully balances individual UAV objectives with global network performance, thus guiding UAVs toward trajectories that simultaneously optimize energy efficiency and network-wide AoI.

\par Additionally, Fig.~\ref{fig4:subfig2} illustrates peak AoI variations corresponding to different values of $\eta_{le}$, where $\eta_{le}$ represents the laser-to-electricity conversion efficiency in Eq.~\eqref{eq:laser}. Notably, almost all algorithms achieve optimal performance at $\eta_{le}$ = 0.15. When $\eta_{le}$ falls below 0.15, the reduced laser charging power extends UAV charging time, limiting their ability to collect IoT information outside the charging zone, thereby increasing AoI. Conversely, higher $\eta_{le}$ values result in shorter charging durations, preventing sufficient data collection within the charging zone, thus also increasing AoI. MAPPO-TM consistently outperforms other algorithms across different $\eta_{le}$ values, achieving a 5-10\% lower peak AoI. This superior performance can be attributed to its LSTM mechanism, which effectively captures temporal dependencies in the UAV charging and information collection processes, thereby enabling more informed decision-making and better system optimization.

\subsubsection{UAV Flight Trajectory Analysis}

\par Fig.~\ref{fig4:subfig3} depicts the flight trajectories of multiple UAVs in the system. The UAVs start near the LBDs with partially charged batteries and initially collect data of IoTs in the charging area. Then UAVs fly out to perform tasks once fully charged, and return to recharge when battery levels drop below a threshold, thereby demonstrating efficient task division with minimal overlap. This optimized behavior is driven by LSTM-based temporal dependency learning of MAPPO-TM, which captures historical trajectories and energy patterns for better long-term planning, and its multi-agent coordination mechanism, which uses a dual-attention approach to balance individual energy management with global AoI minimization. The stable policy updates of the algorithm, facilitated by a clipped surrogate objective and periodic parameter synchronization, further ensure consistent trajectory decisions, thus preventing erratic movements. As a result, MAPPO-TM enhances mission endurance, reduces AoI, and improves overall efficiency.

\par In summary, the proposed MAPPO-TM algorithm consistently outperforms alternative algorithms in terms of cumulative rewards and peak AoI optimization in multi-UAV data collection scenarios. Moreover, the LSTM-based temporal dependency learning enhances decision-making stability, while the dual-attention coordination mechanism significantly improves multi-agent cooperation, thereby resulting in superior overall system performance.

\section{Discussion}

\par In this section, we present some discussions related to the system and algorithm.

\subsection{Scalability Analysis}

\par The considered LBD-powered multi-UAV data collection system is designed with adaptability and flexibility at its core, enabling it to function effectively across various IoT deployments. The following key aspects demonstrate this adaptability:

\begin{itemize}
    \item \textit{The considered system is fundamentally agnostic to the specific hardware characteristics of IoTs, focusing instead on their functional capabilities for data generation and communication.} This design principle allows our approach to work with heterogeneous IoT deployments spanning multiple application domains. While IoTs may vary significantly in their processing power, memory capacity, and energy resources, our system primarily requires them to maintain basic communication capabilities and data buffering functionality. The UAV trajectory planning and laser charging scheduling optimized by MAPPO-TM operates independently of the internal complexities of individual IoTs, instead focusing on their spatial distribution, data generation patterns, and AoI characteristics.
    \item \textit{The computational complexity of our MAPPO-TM algorithm has been carefully analyzed and optimized, as detailed in Section~\ref{sec: MADRL-based approach} of our paper.} During the execution phase, the time complexity reduces to $\mathcal{O}(TN|\theta_a|)$ and the space complexity becomes $\mathcal{O}(N|\theta_a|)$, where $T$ represents the number of time slots, $N$ is the number of UAVs, and $|\theta_a|$ denotes the parameter size of the actor networks. This significant reduction in computational requirements during deployment enables our system to function efficiently even when computational resources are constrained. The LSTM-based temporal memory mechanism in our algorithm is particularly efficient, as it allows for selective retention of only the most relevant historical information, thereby minimizing memory requirements while maximizing decision quality.
    \item \textit{The proposed MAPPO-TM framework employs a CTDE paradigm, which provides substantial deployment flexibility.} The centralized training can be performed on powerful computing infrastructure (e.g., cloud servers or edge computing nodes), while the trained models are deployed to individual UAVs for decentralized execution. This approach effectively separates the computationally intensive training process from the resource-constrained deployment environment. Our simulation results demonstrate that MAPPO-TM maintains stable performance across different laser-to-electricity conversion efficiency values ($\eta_{le}$), highlighting its robustness to variations in the physical characteristics of the deployment environment.
\end{itemize}

\par As such, the MAPPO-TM framework represents a flexible and adaptable solution for AoI optimization in laser-charged UAV-assisted IoT networks. The ability of the MAPPO-TM framework to function effectively across diverse IoT device types, computational environments, and network configurations makes it well-suited for practical deployment in various real-world scenarios.

\subsection{Robustness Analysis}

\par In practical deployments, UAV failures due to hardware malfunctions or environmental factors can significantly impact system performance. Our framework implements specific mechanisms to handle various failure scenarios at a granular level. These mechanisms are designed to address failures of both IoT devices and UAVs during operation, ensuring system resilience through adaptive behavior rather than requiring complete system reconfiguration. The integration of these failure handling techniques directly into our MADRL framework allows for real-time response to unexpected events without compromising overall mission objectives. The details are as follows:

\begin{itemize}
    \item \textit{Faulty IoT Handling:} In our system, when a UAV detects that an IoT is unresponsive or malfunctioning (i.e., no data transmission is occurring despite being within communication range), the UAV updates its observation space $\mathcal{O}_t^j$ to mark this IoT as faulty. This information is then shared with other UAVs during the centralized training phase. The MAPPO-TM algorithm is designed to adapt dynamically to such changes by adjusting the UAV trajectories to prioritize functioning IoTs, thereby maintaining efficient data collection despite IoT failures. Additionally, the temporal memory mechanism in our LSTM-based actor network enables the UAVs to remember which IoTs have previously been identified as faulty, preventing repeated unsuccessful visit attempts in subsequent time slots~\cite{Hahm2021Reliable, Bukhsh2021Energy}.

    \item \textit{UAV Failure During Operation:} Our multi-agent framework has built-in resilience against UAV failures. If a UAV becomes inoperable during a mission, the remaining UAVs can autonomously redistribute their responsibilities through the multi-agent coordination mechanism. Specifically, the dual-attention value function $V_i(s_t) = w_l V^i_{\text{local}}(o^i_t) + w_g V_{\text{global}}(s_t)$ allows UAVs to dynamically adjust their behavior based on global objectives when the system state changes due to a UAV failure. The system detects a UAV failure when it stops communicating its state updates, and the remaining UAVs then update their global value assessment to account for the reduced fleet capacity, resulting in an automatic reallocation of data collection responsibilities among the functioning UAVs~\cite{Xia2022Multi, Sujit2012Multi}.

    \item \textit{Energy-related UAV Failures:} Since energy management is a critical aspect of our system, we have specifically addressed UAV failures related to energy depletion. The penalty term $r_p(t)$ in our reward function (Eq. 10) strongly discourages UAVs from reaching critically low energy states. If a UAV energy level approaches a critical threshold despite these preventive measures, the system initiates a fail-safe protocol: the UAV immediately prioritizes returning to the nearest laser charging area while transmitting its current data payload to neighboring UAVs if possible. This approach ensures that even if a UAV cannot complete its mission due to energy constraints, the collected data is not lost, and the peak AoI performance is preserved as much as possible~\cite{Tahir2023Energy, Xing2023Reliability}.
\end{itemize}

\par Moreover, we can also incorporate some additional robustness measures into the considered system to further improve the ability to resist failures while minimizing the disruption to the data collection service. The specific implementation details of these robustness measures are as follows:

\par \textit{Firstly}, we address the feasible retraining approaches that enable rapid algorithm convergence when UAV failures occur. Specifically, in the event of UAV malfunction, system administrators can adjust the UAV count parameters in the simulation environment and subsequently retrain the DRL algorithm. Given that this process can be executed in computationally robust settings, administrators may proactively develop multiple versions of the DRL algorithm trained with varying UAV quantities as contingency measures. During this retraining process, advanced techniques such as transfer learning~\cite{Weiss2016survey} can significantly accelerate algorithm convergence. Furthermore, by utilizing edge computing infrastructure and implementing incremental learning methodologies~\cite{Zhao2024MEDIA}, we can perform dynamic updates to the deployed neural networks, thereby enhancing UAV adaptability to fluctuations in operational fleet size.

\par \textit{Secondly}, we highlight that redundancy-based fault-tolerance mechanisms offer substantial benefits for such emergency management. For example, maintaining a reserve fleet of standby UAVs allows for prompt deployment to replace malfunctioning units with minimal operational disruption. Such redundancy mechanisms facilitate real-time adaptation and preserve communication system performance without significant service interruptions.

\par These fault-handling mechanisms significantly enhance the robustness of our MAPPO-TM framework in real-world scenarios where hardware failures and energy management challenges are inevitable. The ability to dynamically adapt to such edge cases is a key advantage of our MADRL approach compared to traditional optimization methods, which typically require predefined contingency plans for each possible failure scenario.

\section{Conclusion}
\label{sec: conclusion}

\par This paper has investigated an LBD-powered multi-UAV data collection system in IoT networks. By utilizing LBD to charge UAVs, we effectively improved the flight time of UAVs, thereby enabling the IoT data collection tasks to be completed in a more timely manner. Specifically, we formulated a joint optimization problem that aims to minimize the peak AoI of the IoT network while ensuring that the UAVs follow the charging strategy and do not run out of power, which is characterized by high real-time and dynamic complexity. To address this problem, we proposed a MAPPO-TM algorithm that integrates LSTM with the MAPPO algorithm, which incorporates time-dependent learning to capture the historical trajectory and energy state of the UAV. Furthermore, the algorithm implements a multi-agent coordination mechanism to balance individual and global objectives, thus improving stability and convergence speed. Simulation results validated the effectiveness of the proposed MAPPO-TM algorithm, which demonstrates superior performance compared to other baselines, particularly in enhancing learning stability and reducing peak AoI, and thus highlighting its potential for practical deployment in the LBD-powered multi-UAV data collection system in IoT networks.

\par To further enhance the robustness and applicability of the considered system and proposed algorithm in practical scenarios, future works will focus on the following aspects. Firstly, future works include developing an adaptive fault-tolerance mechanism that can dynamically reconfigure the multi-UAV network topology when failures occur, incorporating rapid retraining methods with transfer learning to minimize disruption to data collection operations. Secondly, we plan to enhance the laser charging efficiency by investigating advanced beam tracking algorithms and optimizing the trade-off between charging time and data collection to further reduce peak AoI in larger-scale deployments. Finally, we will incorporate domain randomization techniques during training to improve the robustness of the MAPPO-TM algorithm against environmental uncertainties and extend its generalization capabilities to diverse operational conditions.

\bibliographystyle{IEEEtran}
\bibliography{cited2.0}
\end{document}